# Clustering of the Diffuse Infrared Light from the COBE DIRBE maps. I. $C(0)$ and limits on the near-IR background.

A. Kashlinsky[1], J. C. Mather[2], S. Odenwald[3], M. G. Hauser[4]

[1] NORDITA, Blegdamsvej 17, DK-2100 Copenhagen, Denmark
[2] Code 685, NASA Goddard Space Flight Center, Greenbelt, MD 20771
[3] Hughes STX Corporation, Code 685.3,
NASA Goddard Space Flight Center, Greenbelt, MD 20771
[4] Code 680, NASA Goddard Space Flight Center, Greenbelt, MD 20771

**ABSTRACT.**

The cosmic infrared background (CIB) and its anisotropy have not yet been measured, but are important signatures of the early evolution and clustering of galaxies. The near IR is particularly interesting because redshift effects bring the peak luminosity of distant galaxies into the near IR, allowing high redshift objects to dominate the theoretical predictions of the CIB and its fluctuations. This paper is devoted to studying the CIB through its correlation properties. We studied the limits on CIB anisotropy in the near IR (1.25, 2.2, and 3.5 $\mu m$, or $J$, $K$, $L$) bands at a scale of 0.7° using the COBE[1] Diffuse Infrared Background Experiment (DIRBE) data. In single bands we obtain the upper limits on the zero-lag correlation signal $C(0) = \langle(\nu\delta I_\nu)^2\rangle < 3.6 \times 10^{-16}$, $5.1 \times 10^{-17}$, $5.7 \times 10^{-18}$ W$^2$m$^{-4}$sr$^{-2}$ for the $J, K, L$ bands respectively. The DIRBE data exhibit a clear color between the various bands with a small dispersion. On the other hand most of the CIB is expected to come from redshifted galaxies and thus should have different color properties. We use this observation to develop a 'color subtraction' method of linear combinations of maps at two different bands. This method is expected to suppress the dominant fluctuations from foreground stars and nearby galaxies, while not reducing (or perhaps even amplifying) the extragalactic contribution to $C(0)$. Applying this technique gives significantly lower and more isotropic limits. For the $J - K$, $J - L$, and $K - L$ combinations these limits are respectively $C(0) < 6.3 \times 10^{-17}$, $1.4 \times 10^{-16}$, $1.2 \times 10^{-17}$ W$^2$m$^{-4}$sr$^{-2}$ respectively. We also use simple no-evolution models to interpret these numbers in terms of the total CIB levels, postponing a more detailed interpretation to a forthcoming paper. From the single band fluctuations we estimate upper limits on the CIB from clustered matter of $(\nu I_\nu)_{z,rms} \equiv \sqrt{\int(\frac{d\nu I_\nu}{dz})^2 dz} < 200$, 78, and 26 nWm$^{-2}$sr$^{-1}$ in the $J$, $K$, and $L$ bands independently of the evolution history or spectral energy distribution. The color-subtracted signals strongly constrain the color evolution of galaxy populations and, if their degree of isotropy is indicative of a cosmological origin, could allow determination of the total diffuse fluxes due to clustered material.

---

[1] The National Aeronautics and Space Administration/Goddard Space Flight Center (NASA/GSFC) is responsible for the design, development, and operation of the *COBE*. Scientific guidance is provided by the *COBE* Science Working Group. GSFC is also responsible for the development of the analysis software and for the production of the mission data sets.

Subjects: Cosmology, Cosmic Background Radiation, Galaxies: Clustering, Galaxies: Evolution

# 1 Introduction

Diffuse background radiation fields contain information about the entire history of the universe, including those periods in which no discrete objects exist or can be detected by telescopic study. The cosmic infrared background (CIB) should contain much of the luminosity of the early generations of stars and galaxies, since the cosmic expansion redshifts the frequency of peak luminosity from these sources from the UV and visible bands into the infrared (IR). In addition, dust absorbs much of the original UV luminosity and re-emits it in the IR. The CIB therefore contains important information about the evolution of the Universe between the redshift of last scattering (probed by the microwave background) and today (probed by optical surveys).

The COBE Diffuse Infrared Background Experiment (DIRBE) was designed for the purpose of searching for the CIB (Boggess *et al.* 1992, Silverberg *et al.* 1993). Though it is hard to distinguish the CIB from foreground sources such as interplanetary and interstellar dust, and stars, extrapolations of galaxy counts clearly predict a near IR CIB well above the DIRBE noise level (Franceschini *et al.* 1991, Pagel 1993). Furthermore, there could have been other sources of the diffuse infrared background such as Population III stars (*e.g.* Kashlinsky and Rees 1983) which would increase the level of the CIB (Bond *et al.* 1986). These diffuse background levels, however, are expected to be significantly below the cumulative fluxes produced by Galactic stars, zodiacal emission, and cirrus. The most direct analysis of the sky brightness using DIRBE data gives the limits on the CIB plotted as crosses in Fig. 1, derived from the darkest parts of the sky (Hauser 1995, 1996). At the $J$, $K$, and $L$ bands they are respectively $\nu I_\nu < 393 \pm 13$, $150 \pm 5$, and $63 \pm 3$ nWm$^{-2}$sr$^{-1}$. The latest attempt by the DIRBE team to account for foreground emission is reported by Hauser (1996), giving residuals for these bands ranging from 50 to 104, 15 to 26, and 15 to 24 nW m$^{-2}$sr$^{-1}$ respectively.

An alternative detection method was suggested by Gunn (1965) based on the correlation properties of the diffuse backgrounds: if the spatial correlation function of the emitters is known, one can derive the diffuse background produced by these emitters from the correlation function of the diffuse background. This is the approach which we follow in this paper. Indeed, if the diffuse background comes from the redshifted light emitted by galaxies, its fluctuations must reflect the correlation properties of galaxies. The CIB might in principle contain contributions from other processes, such as the decay of massive primordial particles, that would not necessarily share the clustering properties of galaxies. This part of the CIB might not be be detectable with the methods discussed here. For brevity, we will ignore this possibility in the rest of this paper, and all of our discussions of the CIB will refer only to the part contributed by matter clustered like galaxies.

Thus the amplitude of the diffuse light correlation function would depend on the amplitude of the correlation function of galaxies, $\xi(r,z)$, their luminosities and spectra and densities as a function of $z$, and the effects of intervening dust on absorption, scattering, and re-emission (Bond, Carr and Hogan 1991). On the relevant scales the galaxy clustering functions are mea-sured quite accurately from the Lick (Groth and Peebles 1977) and APM catalogues (Maddox et al. 1990). Since these functions are known for low $z$ and their evolution can be estimated or modeled, the CIB diffuse light correlation function can be predicted.

The strategy of such an investigation would be the reverse of the previous work on the cosmic microwave background, where the intensity was known and one had to find its correlation properties. For the CIB, the theoretical correlation properties are known from the measured $\xi(r)$, and we aim to find or constrain the background levels. Shectman (1973, 1974) was the first to apply this method in practice in the visible bands. Using photographic plates of an a priori dark part of the sky, he removed bright stars as well as galaxies brighter than $V \simeq 18$ (or $z < 0.1$). The residual extragalactic signal was then separated from the Poisson fluctuations produced by the remaining stars. The amplitude of the $V$-band diffuse background was then found by Shectman to be in good agreement with that measured from galaxy counts (see Peebles 1980, §58). Martin and Bowyer (1989) performed a similar analysis in the far-UV bands ($\lambda = 0.135 - 0.19$ $\mu$m) where the stellar foreground contribution is low. They also reported a correlated extragalactic signal.

The plan of the paper is as follows. In Section 2, we discuss the theoretical expectations. We adjust our models to match the counts observed in deep $K$-band surveys, and confirm the prediction that they should lead to a measurable CIB that is well above the DIRBE noise levels, but substantially below the residual stellar fluxes from the Galaxy. We then predict the amplitude of the diffuse background correlation signal $C(0)$ at zero-lag (the mean square deviation of the map) as seen through the DIRBE beam and show how its value sets an upper limit on the total background flux. In Section 3, we discuss the nature of the various foregrounds that must be suppressed or eliminated to permit measurement of $C(0)$. In Section 4 we present the methods used to analyze the DIRBE $J$, $K$ and $L$-band data, the selection of the sky fields, and the results obtained. We develop a mathematical basis for a foreground subtraction using the color properties of the DIRBE maps in conjunction with the fact that the predicted CIB comes from redshifted galaxies, and show that such analysis can significantly decrease the stellar component of the correlated signal without significantly attenuating the correlated part of the CIB. The interpretation and implications of our findings are discussed in Sections 5-6.

## 2 Predicted IR Correlation Function from Galaxies

### 2.1 Near-IR Background from Normal Galaxies

The expected CIB from galaxy formation and evolution differs in various cosmological models and depends on 1) the evolution of the galaxy spectral energy distribution (SED) with redshift $z$, 2) the evolution of the galaxy luminosity function with redshift, 3) the evolution of galaxy clustering and the power spectrum of the galaxy spatial distribution, and 4) the effects of dust absorption and re-emission. The first three of these are measured quite accurately in the present-day Universe, but little is yet known about the dust, which need not all be local to the luminosity sources. For the purpose of this paper we will ignore the effects of this dust, and defer a proper discussion to a future paper. The present-day SED is known for various galaxy types over the entire range of frequencies from the soft UV to near-IR bands (e.g. Yoshii and Takahara 1988 and references therein), and elaborate predictions can be made from galactic and stellar evolution modelling (e.g. Bruzual 1983). Similarly, the present-day $B$-band luminosity function is known to have the form suggested by Schechter (1976), with the latest compilation of parameters coming from the APM survey (Loveday *et al.* 1992).

We now review the standard approach to predicting the CIB. The bolometric flux received from a galaxy of intrinsic luminosity $L$ at redshift $z$ is $L/4\pi x^2(z)(1+z)^2$, where $x(z) = c\int_0^z (1+z)dt$ is the comoving distance to $z$ and the proper volume of a slice between $z$ and $z+dz$ subtending solid angle $\delta\theta$ is $\frac{dV}{dz} = R_H \frac{x^2(z)}{(1+z)^4\sqrt{1+\Omega z}}\delta\theta$, and $R_H = cH_0^{-1}$. (We consider models with the cosmological constant $\Lambda = 0$). Thus we write the rate of accumulation of the specific intensity of the radiation from galaxies measured today at frequency $\nu$ as:

$$\frac{\partial \nu I_\nu}{\partial z} = \frac{R_H}{4\pi(1+z)^2\sqrt{1+\Omega z}}\mathcal{L}(z)[\nu f_\nu(\nu(1+z);z)], \qquad (1)$$

where $f_\nu \Delta \nu$ is the fraction of the total light observed within a range $\Delta\nu$ wide around $\nu$; we define $f_\nu$ as the SED. The integral of this equation with respect to $z$ gives the total brightness of the CIB. $\mathcal{L}(z)$ is the comoving luminosity density at redshift $z$ and is related to the luminosity function $\Phi$ via $\mathcal{L}(z) = \int \Phi(L;z) L dL$. We ignore for the present the effects of dust absorption and scattering within the galaxies and along the line of sight. Eq (1) must be normalized to the present-day luminosity function in the $B$-band, which obeys the Schechter form $\Phi(L)dL = \Phi_*(L/L_*)^{-\alpha}\exp(-\frac{L}{L_*})dL$, with the present-day value of the luminosity density given by

$$\mathcal{L}(0)f_\nu(\nu_B)\Delta\nu_B = \Phi_*\Gamma(2-\alpha)L_*, \qquad (2)$$

where $\Delta\nu_B$ is the bandwidth of the $B$ band filter. In what follows we adopt the values of the Schechter luminosity function parameters from Loveday *et al.* (1992) which give

$$\Phi_* L_* R_H \Gamma(2-\alpha) = 100 \; nWm^{-2}sr^{-1} \qquad (3)$$

independently of the value of $h$. Note that the number above may be uncertain by a factor of $\sim 2$ due to the uncertainties in the measurements of the galaxy luminosity function from various surveys. If there were no evolution, $\mathcal{L}(z) = \mathcal{L}(0)$. In the present discussion we are interested in observed wavelengths from 1.2 to 3.5 $\mu$m, and source redshifts of a few, so it is enough to know galaxy spectra for $\lambda > 0.5\;\mu$m (see Fig. 2). We adopt the values for the spectra at these wavelengths, together with the galaxy type mixes, from Yoshii and Takahara (1988).

To be consistent with observations, the predictions for the levels of the CIB are based on galaxy distributions tied to the observed counts of galaxies. The data on faint galaxy counts currently cover the range from $B$ to $K$ (2.2 $\mu$m) and contain galaxies typically out to $z \sim 1$. In the $B$-band, the counts go to $B \sim 28$ (Tyson 1988; Lilly *et al.* 1991), in $I$ they reach $I \sim 25$ (Lilly 1993 and references therein), and in $K$ they reach $K \sim 24$ (Cowie *et al.* 1991; Djorgovski *et al.* 1995). The counts in various bands can constrain both the global geometry of the Universe (the $\Omega$ parameter) and the possible amounts of galaxy evolution between now and typically $z \sim 1$ (*e.g.* Koo and Kron 1992; Cole *et al.* 1992). For example, Fig. 1a shows fits to the Cowie *et al.* (1993) $K$ counts (triangles) for dust-free no-evolution models for $\Omega = 1$ (solid line) and $\Omega = 0.1$ (dashed line). The $\Omega = 0.1$ Universe gives a good fit to the data, but if $\Omega = 1$ a certain amount of galaxy evolution would be required (cf. Cole *et al.* 1992, Cowie *et al.* 1993). The amount of this evolution and the value of $\Omega$ are further constrained by the counts in other bands and by the data on the redshift distribution in the samples (Gardner *et al.* 1993; Koo *et al.* 1993).

Fig. 1b shows the $J, K, L$ fluxes based on the $K$-counts data for no-evolution models described in the previous section: triangles correspond to $\Omega = 0.1$ and squares to $\Omega = 1$. The redshift of galaxy formation was taken to be $z_g = 6$ in these calculations, but we find that the numbers are rather insensitive to $z_g$ for $z_g \geq 3$. The reason for this is that at these high redshifts the original frequency is in the UV where the (present-day) SED falls off sharply. The present day galaxy emission is peaked in the visible bands and this is also the reason why the predicted diffuse IR fluxes contain contributions predominantly from galaxies at higher $z$. The amplitude of the background for these models is only weakly dependent on $\Omega$. These simple (no evolution of the luminosity function or galactic SED) estimates show that there should be fluxes above the DIRBE map noise levels (see Table 2 below) in the diffuse IR light due to distant galaxies seen in the faint galaxy surveys (cf. Franceschini *et al.* 1991). Such fluxes carry with them important information about the possible evolution of galaxy populations, the galaxy luminosity function, the conditions in the Universe at high $z$, the intergalactic dust opacity, and the galaxy merger rate (*e.g.* Broadhurst *et al.* 1992). Of course, any realistic galaxy model should predict evolution in both the SED and in the luminosity function. It is possible to construct such models that fit the galaxy counts for any $\Omega$ (*e.g.* Cole *et al.* 1992; Gardner *et al.* 1993; Lilly 1993; Cowie *et al.* 1993). Furthermore, Yoshii (1993) argues that any interpretation of the counts that measure galaxies at such high $z$ must depend strongly on the detection criteria, and also on the magnitude measurement algorithm. This would further complicate models (but see Koo *et al.* 1993 for another way to deduce the evolution from the counts data). It is worth noting however, that the latest $K$ counts from the Keck telescope (Djorgovski *et al.* 1995) show that the counts continue to rise out to $K \simeq 24$ and are described extremely well by the no-evolution low $\Omega$ models.

Here we do not attempt an exact and realistic model for the counts, but only show that galaxies at redshifts in the range out to $z \sim 3$ are capable of providing levels of the near-IR background above the DIRBE noise level. Similarly the predicted fluxes can vary because of the uncertainty in the normalization of (3). For example, choosing the luminosity function from the survey of Efstathiou *et al.* (1983) would result in (3) being a factor of 2.4 higher (cf. eq. 4.7 in Efstathiou *et al.* 1983) leading to correspondingly higher values for the diffuse backgrounds produced by such galaxies. Furthermore, stellar evolution alone would also lead to some changes of the galactic SED. Thus the numbers for the predicted background fluxes in Fig. 1 should be treated mainly for illustrative purposes and order of magnitude comparisons. We therefore follow a different path of interpretation: we will interpret the data on the correlation signal deduced in Section 4 in terms of the cosmological parameters and the total diffuse background levels.

Fig. 2 shows the differential contribution to the total flux in the $J, K, L$ bands per unit of logarithmic redshift interval, $zd(\nu I_\nu)/dz = zd(\lambda I_\lambda)/dz$. Without evolution or source galactic or intergalactic dust opacity, most of the diffuse flux in $J$ band would come from galaxies at $0.2 < z < 1.5$, in $K$ from galaxies at $0.5 < z < 2$, and in $L$ from galaxies at $0.6 < z < 2.5$. As described above this phenomenon results from redshifting the main luminosity of distant galaxies emitted at visible wavelengths into the infrared bands. Presumably, at still longer wavelengths one would be probing galaxies at even larger redshifts, although the information about galactic spectra longward of the $K$ and $L$ bands is not as abundant. Furthermore, at longer wavelengths the luminosity function is better described by the IRAS data (Soifer, Neugebauer and Houck 1987 and references therein, and Saunders, Rowan-Robinson, and Lawrence 1992) rather than the Schechter function. The extrapolation of galaxy spectra to this range is model-dependent, but it is expected that distant galaxies should contribute measurable fluxes for wavelengths at least up to $\simeq 100$ $\mu$m. This is suggested by both the broad-band SEDs observed by IRAS (Hacking and Soifer 1991) and the physical models of a single SED over the entire range of mid- to far-IR (Beichman and Helou 1991). As Fig. 1b shows, the integrated galaxy fluxes in the $J, K, L$ bands are significantly above the DIRBE detector noise levels (Boggess $et\ al.$ 1992), but still significantly below the faintest total sky brightness measured by DIRBE. Our estimates in this paper will show that the correlation method could be helpful in discovering the flux signals from the galaxy populations.

## 2.2 Correlation function of the near-IR diffuse background

As was mentioned in the Introduction, the correlation function of the diffuse background provides information on the CIB if the correlation function of the clustered material emitting the background is known. We quantify this in Secs. 2.2-2.4 to emphasize the importance of measuring the correlation function of the CIB and to show what kind of information it then gives us about the diffuse background.

If the primordial density fluctuations were Gaussian as most theories assume, then the three-dimensional spatial correlation function determines all properties of the primordial density field. The spatial correlation function of galaxies, $\xi(r)$, has now been measured to large distances ($< 100 h^{-1}$Mpc ) in the $B_J$ (Maddox $et\ al.$ 1990; hereafter the 'APM' survey) and $r$ bands (Picard 1991; hereafter the 'Palomar' survey). These estimates are consistent with other measurements of the mass distribution (cf. Kashlinsky 1994 and references therein). While deeper than before, the APM and Palomar surveys map the galaxy distribution at rather low redshifts (the APM survey limit is $b_J = 20.5$, and the Palomar survey limit is $r = 19$).

For deeper samples, the analysis of Efstathiou $et\ al.$ (1991) suggests that Tyson's faint blue galaxies have a smaller correlation amplitude than the APM sample (cf. Bernstein $et\ al.$ 1994). Similar conclusions are derived for a sample of faint galaxies in the $I$-band (Pritchet and Infante 1992). This result could imply that one is now dealing with a new and less clustered population of blue galaxies (cf. Cowie $et\ al.$ 1991), or that clustering has progressed more rapidly in the Universe than the simple models would require (Melott 1992). On the other hand, Cole $et\ al.$ (1994) find from pencil-beam measurements of 1,100 galaxies to a magnitude limit of $B = 22$ that the spatial correlation properties are adequately described by models with little evolution of the correlation length. The galaxy sample identified by IRAS seems to have a smaller correlation scale than found in the optical galaxy surveys, although on large scales the power found in the APM survey in $B$ is consistent with that seen in the QDOT survey of IRAS galaxies (Saunders $et\ al.$ 1991).

The present-day growth factor of galaxy clustering is unity by definition. However the epoch of galaxy formation varies with cosmological scenarios; $e.g.$ in the standard cold-dark-matter (CDM) model normalized to the COBE data on the microwave background anisotropies (Smoot $et\ al.$ 1992), the bulk of galaxy formation would be expected to occur around $z \sim 5 - 6$ (Efstathiou and Rees 1988; Kashlinsky and Jones 1991). For CDM models normalized to the large-scale data on the galaxy distribution (peculiar velocities and large-scale correlations) the epoch of galaxy formation would be significantly more recent (Kashlinsky 1993). In baryon-dominated models in an open Universe, $e.g.$ a primordial isocurvature baryonic (PIB) model (Peebles 1987a,b), the epoch of galaxy formation would occur at larger redshifts than in the standard CDM model. At large redshifts the galaxy SED and the luminosity function must vary due to processes such as galaxy merging, progressive clustering, star formation histories, dust formation and evolution, among others. It is therefore important to constrain the possible evolutionary tracks for these processes. The diffuse radiation field produced by galaxies summed over all redshifts can provide integral constraints on these processes, and shed important light on the epoch of galaxy formation, the history of galaxy evolution, and the value for $\Omega$.

The three-dimensional correlation function is not directly observed in two dimensional images and surveys. We therefore consider the projected two dimensional correlation function as a directly observable quantity. For diffuse light, the projected 2-D correlation function between fluxes in two different bands $\nu_1, \nu_2$ is defined as $C_{\nu_1,\nu_2}(\theta) \equiv \langle \nu_1 \delta I_{\nu_1}(\mathbf{x}) \cdot \nu_2 \delta I_{\nu_2}(\mathbf{x}+\theta) \rangle$, where $\delta I_\nu(\mathbf{x})$ is the fluctuation in the spectral surface brightness $I_\nu$ in the direction $\mathbf{x}$. It is related to the underlying 2-point correlation function via the analog of the Limber (1953) equation: $C_{\nu_1,\nu_2}(\theta) = \int\int \frac{\partial \nu_1 I_{\nu_1}}{\partial z_1} \frac{\partial \nu_2 I_{\nu_2}}{\partial z_2} \xi(r_{12}, z_1, z_2) dz_1 dz_2$, where $r_{12}$ is the proper distance between two galaxies at $z_1, z_2$ respectively that subtend angle $\theta$ (e.g. Peebles 1980). In the limit of small angles $\theta \ll 1$ it becomes

$$C_{\nu_1,\nu_2}(\theta) = \int_0^\infty dz \frac{\partial \nu_1 I_{\nu_1}}{\partial z} \frac{\partial \nu_2 I_{\nu_2}}{\partial z} \int_{-\infty}^\infty d\Delta \xi(r_{12}; z), \qquad (4)$$

where $\Delta \equiv (z_2 - z_1)$ and the proper length is $r_{12}^2 = (c\frac{dt}{dz}\Delta)^2 + \frac{x^2(z)}{(1+z)^2}\theta^2$. Fig. 3a shows the comoving scale subtended by the DIRBE beam as function of redshift. (The proper length would have a maximum at $z \sim 1$ and decrease thereafter as is discussed in the next section). It is apparent from Figs. 2 and 3a that the extragalactic scales probed by the DIRBE beam at zero-lag are $< 20 - 30 h^{-1}$Mpc. Fig. 3b shows the correlation function implied by the APM survey and the COBE microwave background results with the normalization procedure adopted by Kashlinsky (1992). It shows that on the smallest scales accessible to DIRBE, the present day three dimensional correlation function in eq. (4) can be approximated as a power law of slope $\gamma + 1 \simeq 1.7$. Thus as $\theta \to 0$ we model $\xi(r; z)$ as

$$\xi(r;z) = (\frac{r}{r_*})^{-1-\gamma} \Psi^2(z), \qquad (5)$$

where $r_* = 5.5 h^{-1}$Mpc and the function $\Psi(z)$ accounts for the evolution of the clustering pattern, and is defined to be unity for $z = 0$. It can be taken either from simulations (cf. Melott 1992) or from the observations of the evolution of the correlation function with $z$ (Cole et al. 1994). Substituting eq. (5) into (4) leads for small angles to

$$C_{\nu_1,\nu_2}(\theta) = \frac{\Gamma(\frac{1}{2})\Gamma(\frac{\gamma}{2})}{\Gamma(\frac{1+\gamma}{2})}(\frac{r_*}{R_H})^{1+\gamma}\theta^{-\gamma} \int \frac{\partial \nu_1 I_{\nu_1}}{\partial z}\frac{\partial \nu_2 I_{\nu_2}}{\partial z} \Psi^2(z)[\frac{R_H(1+z)}{x(z)}]^\gamma (1+z)^2 \sqrt{1+\Omega z} dz. \qquad (6)$$

Note that there are two important regions of the integrand here. The appearance of $x(z)^{-\gamma} \propto z^{-0.7}$ at small $z$ produces an integrable singularity at $z = 0$, confirming the intuitive expectation that foreground galaxies should contribute a large fluctuation. However, the evolutionary, redshift, and geometrical effects produce another peak in the integrand at high redshift. This is the reason for the importance and potential power of this approach. The color subtraction method we have developed to reduce the influence of foreground stars will also suppress the contribution to the IR correlation from the singularity at $z = 0$; see below.

It is convenient to parametrize the evolution of the luminosity function and of the galaxy SED as

$$E(z) \equiv \frac{\mathcal{L}(z)}{\mathcal{L}(0)}, \tag{7}$$

$$S(\nu;z) = \frac{f_\nu(\nu(1+z);z)}{f_\nu(\nu;0)}. \tag{8}$$

The parameters $E(z)$, $S(\nu;z)$, and $\Psi(z)$ define all the possible types of evolution relevant for the present analysis. By definition, $E(0) = S(\nu;0) = \Psi(0) = 1$. Note that without evolution, $E(z) = 1$, but the spectral function $S(\nu;z)$ would still vary due to redshift effects.

With this notation the intrinsic correlation signal on subdegree scales becomes

$$C(\theta) = 1.5 \times 10^{-3} (\frac{\theta}{1^\circ})^{-\gamma} [\frac{\Gamma(2-\alpha)\Phi_* L_* R_H}{4\pi}]^2 [\frac{\nu f_\nu(\nu;0)}{\Delta\nu_B f_\nu(\nu_B;0)}]^2 \int \frac{[E(z)S(\nu;z)\Psi(z)]^2}{(1+z)^2\sqrt{1+\Omega z}} [\frac{R_H(1+z)}{x(z)}]^\gamma dz. \tag{9}$$

The right-hand-side of this equation depends only on the evolution parameters and does *not* depend on the uncertainties in the large-scale power distribution.

At longer wavelengths, $\lambda > 10\ \mu m$, the analysis would be somewhat different and more model dependent. The luminosity function at these wavelengths is given by IRAS and is significantly different from the Schechter function (Soifer et al. 1987), and galactic spectra are dominated by thermal emission from dust. There are several luminosity evolution models that are consistent with the number counts at 60 $\mu m$ (Hacking et al. 1987; Lonsdale et al. 1990). Gregorich et al. (1995) used deeper IRAS exposures to find number counts at 60 $\mu m$ that are twice as large as previous counts at the same flux level. Hacking and Soifer (1991) have shown that the number counts require measurable diffuse background from galaxies at $\lambda = 25$, 60 and 100 $\mu m$. They point out that strong density evolution would result in the starburst galaxy population exceeding the total galaxy population by $z \sim 0.5$, and this can not continue to high $z$. Wang (1991) and Beichman and Helou (1991) provide further modelling of galaxies and show that there should be an appreciable CIB at these wavelengths. All this makes us optimistic about extending our theoretical analysis to at least $\lambda \leq 100\ \mu m$.

### 2.3 Convolving with DIRBE beam

In realistic measurements one should convolve eq. (6) with the beam profile, which in the case of DIRBE has a square top-hat profile. Of particular interest is then the zero-lag convolved signal, $C_\vartheta(0)$, which is the same as the mean square fluctuation of the map, $\langle(\nu\delta I_\nu)^2\rangle$. For a circular top-hat beam of radius $\vartheta$ the zero-lag convolved signal is related to the intrinsic one via $C_\vartheta(0) = DC(\vartheta)$. For the correlation function given by (6) the factor $D$ is given by:

$$D(\gamma) = \frac{2^{1-\gamma}\Gamma(1-\gamma/2)}{\Gamma(\gamma/2)} \int_0^\infty y^{\gamma-1} W^2(y) dy \tag{10}$$

where $W(y) = [2J_1(y)/y]$. For $\gamma = 0.6, 0.7, 0.8$ the convolution factor is $D(\gamma) = 1.3, 1.4, 1.5$ respectively and for $\gamma \geq 0.3$ can be approximated as $D(\gamma) \simeq 1.8/(2-\gamma)$.

The nominal square DIRBE beam has the same rms radius as a circle with a radius of $0.4^\circ$. The effective beam is increased by 1) the finite exposure time during the motion of the satellite,

2) the pixelization of the samples, and 3) the errors in the attitude solutions. These factors increase the effective rms beam radius by $\simeq 15\%$, leading to $\vartheta \simeq 0.46°$.

If we use the parameters of eq. (3), the spectral characteristics from Yoshii and Takahara (1988), $\gamma = 0.7$ and $\vartheta = 0.46°$, the expected zero-lag signal in the DIRBE maps becomes:

$$C_\vartheta(0) = 2.2 \times 10^{-19} Q_\nu^2 \int \frac{[E(z)S(\nu;z)\Psi(z)]^2}{(1+z)^2\sqrt{1+\Omega z}}[\frac{R_H(1+z)}{x(z)}]^{0.7}dz \frac{W^2}{m^4 sr^2}, \quad (11)$$

where $Q_\nu = \frac{\nu f_\nu(\nu;0)}{\Delta\nu_B f_\nu(\nu_B;0)}$ and is $Q_J = 8.9$, $Q_K = 3.8$, $Q_L = 1.7$ for the $J, K, L$ bands. If there were no evolution, the numerical value for the integral on the right-hand side of (11) would be $\simeq 2, 4$ and $4.5$ for the $J, K, L$ bands respectively, with no noticeable dependence on either $\Omega$ or $z_g(> 3)$. Note that the analysis of deep counts by Koo et al. (1993) suggests that there was little or no evolution. This is also supported by the recent finding of apparently normal elliptical galaxies with old stellar populations at high ($z \sim 2.5$) redshift (Hu and Ridgway 1994). Driver et al. (1995) used deep HST exposures to find that there is significant evolution of faint blue irregular galaxies. However, infrared fluxes are much less sensitive to recent star formation than are the blue fluxes.

Thus we expect the zero-lag extragalactic signal for the DIRBE maps to be near $C_\vartheta(0) \sim 10^{-2}(\nu I_\nu)^2$ (see sec. 2.4). If the predicted levels of near-IR fluxes change, the numbers for $C_\vartheta(0)$ require corresponding rescaling by a factor roughly proportional to the (total flux)$^2$. Alternatively, one can use the limits on $C(0)$ from the DIRBE maps to infer information on the total levels of CIB, as will be done later in this paper.

In what follows we will drop the subscripts $\nu$ and $\vartheta$ and refer only to the beam-convolved, zero-lag signal.

## 2.4 Zero-lag correlation of the DIRBE sky and the total background

The results of the previous section, if combined with equations relating the fluctuations of the CIB to the total brightness, would allow us to set constraints on the total diffuse background in the DIRBE $J, K, L$ bands via the measurements of the zero-lag signal from the DIRBE maps. Strictly speaking such constraints would be valid only for the clustered component of the CIB. However, except for exotic sources like decaying elementary particles, it is reasonable to assume that any form of matter emitting near-IR radiation does so because localized (peculiar) gravity fields caused it to collapse. It must therefore be clustered and reflect the distribution of the peculiar density field. For the following discussion we consider only the clustered component of the CIB, and note that we are sensitive to any correlated emissions, not only those from discrete galaxies.

To see how the measurements of the correlation signals constrain the levels of the background it is illustrative to rewrite (11) as:

$$C(0) = 3.6 \times 10^{-3} \int (\frac{d\nu I_\nu}{dt})^2 \Psi^2(t)(\frac{R_H}{r})^\gamma H_0^{-1} dt \quad (12)$$

where $t$ is the cosmic time and $r = x(z)/(1+z)$. In the above we adopted $\vartheta = 0.46°$ and $D = 1.4$, which correspond to the DIRBE beam size and nearly top-hat profile. Note that the fluctuation estimate is intrinsically second order and hence is sensitive to the variations in rate of production of the CIB. Since this is not known from observations, a theoretical model of the evolution of the CIB is required to connect the fluctuations with the total. A similar equation can be written for the color subtracted zero-lag signal defined later in the paper. One can see that any measurement of the zero-lag signal should set an integrated constraint on the total squared rate of growth of the flux, defined here as $\int (d\nu I_\nu/dz)^2 dz$. Because of the factor $(R_H/r)^\gamma$ which increases towards modern times ($z \to 0$) and the factor $\Psi^2$ which, if anything, decreases towards early epochs, such flux measures will be more dominated by the emission from the nearby galaxies than the corresponding total flux if determined directly. In the Friedman-Robertson-Walker Universe the ratio of $[r(z)/R_H]^\gamma$ reaches a maximum of 0.45 at $z \sim 1.5$. The value at the maximum is almost independent of $\Omega$, being 0.43 for $\Omega = 0.1$ and 0.47 for $\Omega = 1$. This translates into $[R_H/r(z)]^\gamma \geq 2.2$. Thus any detection of the cosmological $C(0)$ or an upper limit on it from the DIRBE maps would set an <u>upper</u> limit on the total flux:

$$C(0) \geq 8 \times 10^{-3} \int (\frac{d\nu I_\nu}{dt})^2 \Psi^2(t) H_0^{-1} dt \tag{13}$$

One can see that the integral on right-hand-side of (12) represents essentially a weighted squared derivative of the flux with respect to cosmic time emitted by clustered material in frequency band $\nu$. In particular, it shows that if there were a sudden burst of emission at a particular time, the derivatives would be high and the spatial fluctuations as they were at that moment would dominate the observed $C(0)$. Indeed, one can rewrite $\int (\frac{d\nu I_\nu}{dt})^2 \Psi^2(t) H_0^{-1} dt = \int (\frac{d\nu I_\nu}{dz})^2 [\Psi^2(z)(1+z)^2\sqrt{1+\Omega z}] dz$. The growth factor is generally expected to vary as $\Psi^2(z) \propto (1+z)^{-(3+\epsilon)}$. The (extreme) case of low evolution is $\epsilon = -2 + \gamma$, or $\Psi^2(z) \propto (1+z)^{-1.7}$, the case of "painted-on" clustering expected for strong biasing when clustering is expected to be stable in comoving coordinates. The case of high evolution would correspond to the clustering pattern stable in proper coordinates, or $\epsilon = 0$. Thus, in general, the product $[\Psi^2(z)(1+z)^2\sqrt{1+\Omega z}]$ should have only weak dependence on $z$ over the relevant range of redshifts. In other words the total CIB flux for the DIRBE sky should always be:

$$(\nu I_\nu)_{z,rms} \equiv \sqrt{\int (\frac{d\nu I_\nu}{dz})^2 dz} \leq 11\sqrt{C(0)} \tag{14}$$

Thus apart from other benefits, this method allows for a simple, and as we show below efficient, way of constraining the total CIB from the measurements of variance (zero-lag) in the DIRBE maps.

Strictly speaking, the flux measure constrained by (14) differs slightly from the usually defined isotropic flux $\int \frac{d\nu I_\nu}{dz} dz$, in a way that depends on the details of the evolution. However, for the no-evolution models with spectra from Yoshii and Takahara (1988) the two definitions differ by only $\sim 10\%$. In any case a given spectral or evolution history determines (14) just as uniquely as the usual mean fluxes. Furthermore, the comparison between the two fluxes would provide a constraint on the rate of flux emission with redshift.

In the conventional DIRBE analysis one derives upper limits from the DC component of the signal; as eq. (14) shows the correlations method could be more efficient when the foreground emission is very bright but homogeneous. This is typically the case in wavebands longward of the ones considered here where zodiacal light dominates stellar emission. The analysis at these wavelengths will be presented in a future paper. Nevertheless, applying the techniques in the near-IR bands already gives promising and interesting results.

# 3 Foreground Characterization and Removal

The primary mission of the DIRBE was to find or set very strict limits on the CIB contribution from primordial galaxies and other distant sources, and the calibrated data have now been made public. However, the brightness of the CIB remains difficult to determine for a very good reason. The detection and measurement of the CIB is limited by our ability to distinguish it from a variety of non-cosmological foregrounds (Hauser 1993, Wright 1992). In the DIRBE $J$, $K$ and $L$-bands, three foregrounds prevent any simple extraction of the cosmological 'signal' from the all-sky data: 1) scattered zodiacal light and emission from dust within the solar system, 2) Galactic emission due to stars and other compact sources, and 3) emission, absorption, or scattered light correlated with interstellar dust within the Galaxy (the infrared 'cirrus', Low et al. 1984).

Initial results have been published by Hauser (1995, 1996) based on the brightness of the darkest parts of the sky observed by DIRBE, but these conservative limits test only a few exotic theories of relevance to galaxy evolution and cosmology. Less conservative estimates were also presented and revised by Hauser (1995, 1996), based on an initial subtraction of the foreground zodiacal and Galactic emission. The residuals of the fits are of the order of (20-40)% of the total sky brightness, though these are not claimed as CIB limits. Further work might reduce these residuals or reveal the CIB directly.

Indirect limits or a possible detection of the CIB have also been obtained by analysis of the gamma ray spectrum of the BL Lac object Mkn 421. The production of electron-positron pairs from the interaction of the gamma rays with the photons of the infrared background limits the mean free path of the gamma rays. Results have been published by Stecker and DeJager (1993), Dwek and Slavin (1994), and Biller et al. (1995). The limits are not especially low in the near IR, but in the mid IR around 10 $\mu$m where the zodiacal light prevents direct observation, the limit may actually be a detection. Unfortunately the unattenuated source spectrum is not known, so the interpretation is intrinsically model dependent.

A measurement of $C(0)$ must satisfy three criteria to be considered a firm CIB detection. First, it must be isotropic to a high degree. Second, it must have a level significantly above the random noise levels from the instrument. The spatial power spectrum of the fluctuations must be shown to arise outside the instrument and data reduction process. Third, it must be clear that it does not arise from foreground sources or the processes used to remove them. It must not itself be spatially correlated with foreground sources like the Galactic or ecliptic plane, or the local supercluster, since in this case we are interested in the high redshift part of the distribution. In this paper we do not attempt to prove that all these criteria are met. Thus, our results must be regarded as upper limits. At any rate even as such they already provide interesting tests of evolution and clustering processes through eqs. (11) and (14).

## 3.1 Scattered Zodiacal Light

Zodiacal light from solar radiation scattered and emitted by interplanetary dust is an important source of diffuse foreground emission in the near-IR bands. Semi-empirical models of the zodiacal light have recently been developed using the DIRBE data (Hauser 1995, 1996, Reach et al. 1996). These models reproduce the observed zodiacal emission and its variation as a function of time, solar elongation, and frequency, to within a few percent of the total sky emission at these frequencies. Near the ecliptic plane, at latitudes < 20°, the asteroidal dust bands originally detected by IRAS contribute foreground structure of a few percent of the total zodiacal light at angular scales of a few degrees. Outside the ecliptic plane, however, small scale zodiacal structure is of small amplitude (less than a few percent), and the zodiacal light is well represented by spatially smooth models that vary slowly with time. There are some exceptions. Reach *et al.* (1996) found that there is evidence of dust density enhancements caused by low order orbital resonances with the Earth, as suggested by Dermott *et al.* (1994), with a ring and blobs at 1.05 AU. Comet dust trails identified in the IRAS data by Sykes and Walker (1992) are not evident in the DIRBE data, due at least in part to the large beam size. The four bright comets appearing during the DIRBE survey have been excised from the maps used here. For studies of the CIB correlation function, this implies that at large enough ecliptic latitudes, the removal of a simple smooth zodiacal foreground can be supported on the basis of the current generation of zodiacal light models.

## 3.2 Stars

The Galactic foreground $J$, $K$, and $L$-band emission is entirely dominated by stars. At high Galactic latitudes, many stars are clearly recognizable as small groups of bright pixels, but even so there is a strong diffuse flux from a distribution of many faint stars. For purposes of measuring the CIB, regions of the sky within $(20-30)°$ of the Galactic plane can be ignored. The foreground contribution from individual high latitude bright stars, however, cannot be completely removed since at the resolution of the DIRBE beamwidth, all beams include such stars.

The zero-lag signal $C(0)$ is an average over $N_{obs}$ independent beams and can be determined to a fractional precision of order $1/\sqrt{N_{obs}}$, depending on the shape of the distribution of values, and on how many pixels must be eliminated due to bright stars or other point sources. The number of observations $N_{obs}$ is about 1/5 of the number of pixels because the pixels are smaller than the beamwidth. The full sky contains about 80,000 beamwidths.

## 3.3 Infrared Cirrus

According to the IRAS survey, much of the diffuse mid- and far-IR emission at high Galactic latitudes originates in infrared 'cirrus' associated with interstellar dust, which is generally correlated with HI column density (Low *et al.* 1984, Boulanger *et al.* 1985). Considerable dust is present even at the Galactic poles (Hauser *et al.* 1984) and roughly follows a cosecant law in Galactic latitude. At the poles, the $V$ band extinction is $A_v = 0.09 - 0.13$. For latitudes in excess of 20°, the predicted $V$-band extinction obtained from the cosecant law is $A_v < 0.3^m$. At $J$-band we estimate $A_J < 0.1^m$, so extragalactic fluxes could be attenuated by up to 10% at $J$-band.

The present paper is the first in a series and makes no adjustments for the contributions of the infrared cirrus, so our results are necessarily upper limits. It is nevertheless worthwhile to discuss the effects that cirrus could produce. The cirrus emission was found in emission in the IRAS data from 12 to 100 $\mu m$ by Low *et al.* (1984), and indeed is seen with high contrast at 100 $\mu m$. It had previously been seen in reflection by photographic techniques. It is also seen in absorption by its modulation of visible wavelength galaxy counts, by its reddening of individual stars and distant galaxies, and for some lines of sight by narrow line features in stellar spectra.

It is known to be correlated with H I and CO. It has been seen in emission in the near IR by the DIRBE team (Weiland *et al.* 1995, Bernard *et al.* 1994). Clearly the observational approach is to correlate the near IR maps with other indicators of the presence of cirrus. If the correlation is negative (as it is for galaxy counts) then we would deduce the presence of an external CIB that is attenuated by the dust, while if it is positive we would conclude that cirrus emission or scattering is important. Care is needed because in principle there could be such a correlation in the properties of external galaxies themselves. The proof that an observed fluctuation is extragalactic would be nontrivial.

# 4 Data Analysis

The DIRBE was described by Boggess *et al.* (1992) and Silverberg *et al.* (1993). The instrument is a 10-band photometer system covering the wavelength range from 1.25 to 240 $\mu$m with an instantaneous field of view of $0.7° \times 0.7°$. Accurate beam profile and beam response maps were obtained in flight by observing multiple transits of bright point sources. All spectral channels simultaneously observe the same field-of-view on the sky. Additional details of the instrument and data processing are explained in the COBE DIRBE Explanatory Supplement (1995).

## 4.1 Photometric Accuracy and Instrumental Noise

Absolute photometry was obtained through 32 Hz chopping between the sky and an internal zero flux surface maintained at temperatures below 2.5 K. System response was monitored every 20 minutes throughout the 10 month mission by observing an internal thermal reference source. The DIRBE was designed to achieve a zero-point to the photometric scale with an uncertainty below $\nu I_\nu = 1\ nWm^{-2}sr^{-1}$ at all frequencies. Tests conducted prior to launch and in flight indicate that this goal was achieved at the $J, K, L$ bands.

Long-term photometric consistency was established by monitoring a network of stable celestial sources. The primary absolute photometric calibrator for the $J$, $K$ and $L$-bands was the star Sirius. The DIRBE photometry bands are similar in bandpass to the standard Johnson photometric system but are slightly different, and all references to DIRBE photometry use the DIRBE system, not the corresponding Johnson system. Long-term stability of the instrumental photometric scale at the $J$, $K$, and $L$ bands is better than 1%. The long-term stability of the band-to-band intensity ratios, or colors, is 1.4%. The absolute photometric uncertainty in the DIRBE $J$, $K$ and $L$-bands is $\approx 4\%$, and the band-to-band colors are good to $\approx 4\%$. Since DIRBE is a broad-band instrument, the measured in-band intensities are reported as spectral intensities at the nominal frequencies of each band, using a bandwidth appropriate for a source with a flat $\nu I_\nu$ spectrum. Sources that have an SED significantly different from a flat $\nu I_\nu$ will require color corrections. In the $J$, $K$ and $L$-bands, these color corrections are of the order of a few percent or less for blackbodies over the temperature range from 1500 to 20,000 K, which should be representative of redshifted galaxies out to redshifts of 3. We have made no color corrections for this analysis. Additional details on the DIRBE photometric system may be found in Hauser (1993) and the COBE DIRBE Explanatory Supplement (1995).

The DIRBE suffers from a variety of instrument-related noise sources: short and long-term gain fluctuations, cosmic ray hits on the detectors (glitches), digitization roundoff and non-linearities, Johnson noise in load resistors, preamp JFET voltage noise, and photon quantum statistics. The total noise associated with all these sources may be determined from the residuals of the fits of signal versus time for each pixel. Each pixel is observed several times each day for a period of several months, so that there are of order 200-500 observations for each, with nearly an order of magnitude more samples available towards the ecliptic poles.

At these short wavelengths where stars are bright and visible in many pixels, small errors in instrument attitude information are important. Observations whose beam centers fall within a given map pixel are grouped together as though they were identical observations, but they are not. The pixel area is about 1/5 of the beam area, so we would expect an intra-pixel variance for single observations that is of order 1/5 of the inter-pixel variance. All bands are observed nearly simultaneously and with nearly the same field of view. If this design requirement were perfectly satisfied, then the observed colors would be correct even in the presence of pointing variations. As we will use the colors to help suppress foreground emissions, this feature is important.

## 4.2 Maps

The DIRBE data are pixelized using the quadrilateralized spherical cube as described by Chan and O'Neill (1974) and Chan and Laubscher (1976). The transformation from ecliptic to pixel coordinates is straightforward, and has been discussed by White and Mather (1991) as well as in the COBE explanatory supplements (COBE, 1995b). In this representation, the entire sky is projected onto the six faces of a cube, each divided into a square array of $2^n \times 2^n$ pixels. The mapping is chosen so that all pixels have the same solid angle on the sky, so the Jacobian of the transformation is a constant. Solid angle integrals transform neatly into sums over pixels, and shape distortions are minimized. The worst shape distortions are at the cube corners, where a pixel is approximately a rhombus with 60° and 120° angles. Our selected fields avoid these corners, and shape distortion is not important to the computation of $C(0)$ in any case.

At the first stage of analysis we chose four fields located outside the Galactic plane, $|b^{II}| > 20°$. Each was a contiguous field of 128×128 pixels, and covers 38.5° × 38.5°. Two fields (in Faces 1 and 3 of the cube) are located in the Ecliptic plane and two (Faces 0 and 5) near the north and south Ecliptic poles.

In order to compare directly the results of our method with the standard DIRBE program of limiting the background we also analyzed the same five fields that are used in setting the strongest current limits on CIB (Hauser 1995, 1996): four fields are 32 by 32 pixels in size (10° × 10°) and are centered on the two ecliptic (hereafter referred to as SEP and NEP) and two galactic (SGP and NGP) poles. The fifth field is 5° by 5° and is centered on the Lockman hole (hereafter LH), the region of the minimum HI column density at $(l^{II}, b^{II}) = (148°, +53°)$ (Lockman et al. 1986). One of these fields (SEP) contains the Large Magellanic Cloud (LMC) and the other (NEP) has a substantial fraction of pixels lying very close to the Galactic disk $(b^{II} < 25°)$. Thus we also analyzed two fields of the same size but shifted away from the LMC (hereafter SEP1) and the Galactic disk (NEP1).

In order to check the consistency of our results we have studied two different maps, A and B. To create Map A, we chose subsets of the DIRBE data centered on the time when the field was 90° from the Sun. The standard DIRBE weekly sky maps provide intensities averaged over a whole week of observations at a variety of solar elongations. For a pixel at the ecliptic equator, the weekly-averaged elongations can range from roughly 64° to 124°. For each pixel, the intensity at exactly 90° elongation was determined by fitting a smooth function to the nearest four weekly-averaged intensities. No model of the zodiacal light was subtracted from map A.

Map B was derived from the entire 41 week DIRBE data set, now available from the National Space Science Data Center. Note that some fields in map B were not observed for all 41 weeks. A parametrized zodiacal light model developed by the DIRBE team (Reach *et al.* 1996) was used to remove the zodiacal light from each weekly map. The model uses the time dependence of the measured sky brightness as the unique signature of the zodiacal light. The model has 28 adjustable parameters to represent the brightness at all 10 wavelengths, and all wavelengths are fit simultaneously with the same distribution of dust density. Its accuracy was tested by comparing the residual maps obtained at different times during the year; they should be consistent since the time dependence was the signature of the zodiacal light.

We now describe two steps that were used and developed to eliminate the effects of the zodiacal and stellar foregrounds: discrete source masking, and color subtraction. While most of our results will be given as $\lambda I_\lambda = \nu I_\nu$ in units of $nW m^{-2} sr^{-1}$, some plots will be shown in the original $I_\nu$ map units of MJy/sr. The conversion factors are 1MJy/sr = 2400, 1363 and 857 $nW m^{-2} sr^{-1}$ for the $J, K, L$ bands respectively.

## 4.3 Step 1: Discrete Source Masking

Fig. 4 is a surface plot of the brightness for the map A Face 0 field at $J$ band. The appearance of all the fields is similar: a large number of bright point sources situated on a smooth distribution. Direct comparison with the SAO catalog shows that the point sources are virtually all cataloged stars. For purposes of estimating the properties of the CIB we must eliminate the contribution from the discrete foreground as well as possible.

Each field was processed by an iterative background modelling and source removal program developed by the DIRBE team. The smooth component was modeled as a 4th-order polynomial surface, which was fit to the entire field after application of a 5 × 5-pixel spatial filter. Pixels with fluxes greater than $N_{cut}$ standard deviations of the input map above the fitted background model were blanked, as were the surrounding 8 pixels. In other words, each star was assumed to occupy a 3 × 3-pixel region 0.97° square, a factor of ∼ 2 larger in area than the nominal DIRBE beam size of 0.7° square. Some bright stars have images detectably larger than 3 × 3 pixels, since the observations are not all centered in a given pixel, the attitude information for the spacecraft is not perfect, the pixels are not really square but are rhomboid because of the map projection, and the DIRBE beam is smeared on the sky by the scanning motion of 0.3° in the 1/8 second per observation. The most statistically significant parts of these tails are removed by successive passes of the iterative algorithm, as follows.

The procedure was repeated for the remaining non-blanked pixels in the map until no further points were masked by the algorithm. The background model was recomputed for the non-blanked pixels in the field, and the blanking process performed again at the same clipping threshold factor, $N_{cut}$. Since the clipping threshold is defined by a factor relative to the standard deviation of the residuals of the remaining pixels, it changes with each iteration and is slightly different for different values of $N_{cut}$. This process was iterated until no new peaks were identified in the current map iteration. Because this process does not actually subtract a background from the data at each step in the iteration, the final product is a sky map containing all of the original low spatial frequency background information, but in which bright point sources have been deleted by blanking their pixels. We tested cutoff thresholds of $N_{cut}$ = 3.5, 5, and 7, and the results are tabulated in Table 1. Most of the variance comes from the few bright, nearby stars, the ones recognizable as discrete peaks in the maps.

After the masks were created individually for all three bands, a combined mask was made, so that all the pixels used for the analysis were acceptable at all three bands. This step guarantees that the data are consistent for all the bands and for the color-subtracted analysis to be described below.

Figure 5 shows the histogram of the fluctuations in the pixel fluxes from the map mean, $\delta I$, for the fields in Faces 0 and 3 after applying the star removal procedure for both map A (Fig. 5a) and map B (Fig. 5b). Histograms from the smaller regions are similar but noisier because there are far fewer pixels. Solid lines correspond to the $N_{cut} = 7$ clipping threshold, short dashes to $N_{cut} = 5$ and the long-dashed lines to the $N_{cut} = 3.5$ threshold. The horizontal axis for each small plot is the deviation of intensity measured in units of $\sigma_7$, the standard deviation of the residuals of the map for that field and frequency made with $N_{cut} = 7$. As noted above, the clipping algorithm derives its $\sigma$ from the remaining pixels, so it is not the same for all values of $N_{cut}$. For map A from which the zodiacal light component has not been removed, the fields in Faces 0 and 5, both located near ecliptic poles, have similar distributions, and fields 1 and 3 located in the ecliptic plane are similar. The differences in the distributions for fields located in the ecliptic plane (faces 1, 3) and those near the poles (faces 0, 5) are significant. We attribute this to the differences in the relative fractions of the zodiacal light and starlight contributed in the four fields. On the other hand, as Fig. 5b shows, after zodiacal light has been subtracted (Map B) all four fields have similar distributions, confirming our interpretation of the differences between regions seen in Map A. All four fields are at high galactic latitudes, centered at $\pm 50°$ and at $+70°$, so the stellar distributions are expected to be similar. Figure 5 shows that the distribution is heavily skewed, with many more bright pixels than faint ones. The skewness is especially clear for the $N_{cut} = 7$ threshold (solid lines). This is expected for resolved individual point sources uniformly distributed in space, which should have $dN/d\delta I \propto \delta I^{-5/2}$. For unresolved, overlapping point sources, this simple power law distribution no longer holds. For this model we should find a ratio of $2^{5/2}$ between the histograms at 2 and 4 standard deviations, and this is approximately true. Elementary simulations show that the general shape of the histograms, including the width of the peak, the steep drop at negative deviations, and the long tail at large positive deviations, is quite consistent with the fluctuations of point sources uniformly distributed in space. Since the SAO catalog shows that point sources at this flux level are predominantly Galactic stars, we expect that even after our method of point source removal, the fluctuations are still due predominantly, if not exclusively, to stars in our Galaxy.

The corresponding histograms for one of our final regions (SGP) are plotted in the top row of Fig. 13. The figures for single bands show that the number of pixels decreases very rapidly with the amplitude of the deviation $|\delta/\sigma_7|$. The measure of the width of the central part of the histogram is not greatly affected by the skewness. We will return to this point in Section 4.4. The mean flux of the remaining pixels in each field does not vary appreciably with the clipping threshold. On the other hand, the width of the histograms decreases substantially as one goes to lower clipping thresholds. Since the width is a measure of the zero-lag signal, $C(0)$, the results in the figures in conjunction with eq. (14) show that the correlation method can indeed be an efficient tool for studying CIB levels. Fig. 6 shows the fraction of pixels available for the analysis as function of $N_{cut}$ for the five final regions. While it may be tempting to use lower values of the clipping threshold, thereby decreasing the width of the histograms even further, the value of $N_{cut} = 3.5$ probably represents the lowest threshold where enough pixels still remain for analysis. As noted above, the algorithm removes 8 neighbor pixels for every one above the threshold. There might also be some risk of removing genuine CIB fluctuations if the cut is too deep; we have only verified that the SAO star catalog matches our masks down to the $N_{cut} = 7$ level. On the other hand, for realistic levels of the CIB one expects the extragalactic fluctuations removed at the $N_{cut} = 3.5$ level to be very high peaks of the CIB field so that they do not contribute significantly to the total extragalactic $C(0)$.

Inspection of the residual maps shows that there are large-scale gradients present which are produced by the foreground emission from the Galaxy and the Solar system. Since these gradients are clearly not due to a cosmological signal, we removed them prior to computing $C(0)$. Only linear gradients were removed in this process. The reason for limiting ourselves to only first order terms was to not alter significantly the power spectrum of any cosmological signal that may be present; on small scales such a signal should have a power spectrum proportional to (wavenumber)$^{-1.3}$. The number 1.3 comes from the integral transform, which shows that a correlation to $\xi(r) \propto r^{-\gamma-1}$ leads to a power spectrum $P(k) \propto k^{\gamma-2}$. As for the underlying cosmological $C(0)$ we expect that it, being dominated by the small scale structure of the CIB, will remain relatively unchanged by removal of large-scale gradients. We find that the removal of gradients in the $J, K, L$ bands makes only a minor difference to the zero-lag signal for all regions except the SEP field. The SEP field contains a bright and extended feature (LMC) which cannot be identified in the star removing algorithm that we used.

Table 1 shows the values of $C(0)$ in the $J$, $K$ and $L$-bands for map B computed for each of the seven fields using different thresholds for removing bright stars and after the linear gradient removal. Column 1 gives the location of the field. Columns 2 and 3 give the clipping threshold $N_{cut}$ for star removal, and the number of pixels left after the bright point source removal procedure. Columns 4-6 show the value of the residual zero-lag signal in the $J, K, L$ bands respectively, and it is clear that $C(0)$ determined at this stage in the analysis cannot be ascribed to the extragalactic background. It is very anisotropic, with variations by a factor of $> 10$, and depends strongly on $N_{cut}$. The SEP region, which contains the Large Magellanic Cloud, is particularly outstanding. After inspecting the residual maps we attribute this signal to the presence of the LMC. On the other hand, the field SEP1 has a significantly lower signal. These results, while clearly still dominated by the foreground emission, can set constraints on possible galaxy evolutionary tracks, and these will be discussed in Sec. 6. We conclude from this analysis that, in order to measure the predicted underlying extragalactic signal, we must reduce the stellar and interstellar contribution to $C(0)$ by a factor $> 10$ in $J$, $> 5$ in $K$ and $> 2$ in $L$ bands.

Table 2 gives estimates for the contribution of observational noise to these measures of $C(0)$. We evaluated these noise levels for two values of $N_{cut}$ of 7 and 3.5, and for three different time scales, and scaled all of them to account for the different numbers of observations. The data are given for two fields NEP and NGP. The average number of weeks during which the pixels of the region were observed is given in the third column of the table. For each mask three difference maps were made to determine the noise levels on different time scales. Map C was made from 8 weekly maps of each region, after subtracting a model for the zodiacal light. They were combined with alternating + and − signs. Map D was similar, but the comparison was between 4 weeks in the early mission and 4 more weeks six months later, so this measure is sensitive to long term trends in the instrument or in the zodiacal light and its modeling. Map E was similar but was a comparison of two successive weeks. The numbers decrease substantially with $N_{cut}$. The residual variance diminishes roughly inversely with the weight of the observations, as expected if random short-term errors are the dominant noise source. After accounting for that factor, the residual noise seems greater when determined over longer time intervals. In summary, Table 2 gives the values of noise that should be seen in Map B, based on extrapolations of the noises observed in the other maps using the number of weeks of observation as an estimator for the statistical weight of the data. This is still an approximation because each pixel is observed a different number of times in each week.

To investigate the cause of the observational noise we examined the difference maps directly. We found that the difference maps had greatest signal levels in the vicinity of bright stars, as would be expected if the noise comes from errors in assigning the observations to the pixel centers. This is true even after blanking the bright stars; it appears that the mask derived from the average map is not perfect for the individual weeks, as though the beam profile or the attitude errors are slightly different from week to week. The residual noise level also appears to be significantly anisotropic on the sky. These differences should be greatest over long time scales because the DIRBE scan direction and beam smearing are quite different for observations separated widely in time, so Map D should have the largest effective noise level. We also made angular power spectra by Fourier transformation of Maps A-E. Maps A and B showed the clear signature of discrete stars, with a roughly flat noise level for angular scales greater than the beamwidth, and a reduced response for scales that are too small to be resolved by the beam. The difference maps C-E showed the same signature at a reduced level, confirming that these errors are external to the instrument itself. It is clear that the single-band $C(0)$ is not the result of instrument noise.

By contrast, we will show below that the color-subtraction method reduces the random noise terms as well as the foreground star fluctuations. For observations taken only a week apart, the angular power spectrum of the color-subtracted noise appears to be flat for angular scales less than $q^{-1} = 1°$, where $q$ is the angular frequency in radians/degree. There is little sign of the beam profile in these noise power spectra, confirming that the color subtraction also reduces the observational noise originating in attitude errors, the beam profile, and the pixelization process. Nevertheless, the color-subtracted maps from the B data set have a residual variance that is roughly isotropic and above that expected from the short-term observational noise shown in Table 2.

## 4.4 Step 2: Color Subtraction

The signal from the foreground radiation may be disentangled from that of galaxies by considering the properties of the foreground stellar emission that are different from the CIB. The extragalactic objects have somewhat different colors than the Galactic stars, and they are redshifted as well. As Fig. 2 indicates the latter could be a significant effect. Furthermore, galaxies of different morphological types have different stellar populations and this too should help distinguish the colors of the foreground emission from our Galaxy (type Sc) from those of more representative galaxies (E/S0, Sa, etc.) which are expected to dominate the CIB at $J$, $K$ and $L$-bands.

The bottom three boxes in Fig. 7 plot the scatter diagram for fluxes in the DIRBE pixels (in MJy/sr) for the NEP field after the removal of discrete point sources at the $N_{cut} = 5$ threshold for map B. Other fields and map A have similar distributions. There is a clear correlation between the fluxes in various bands indicating the presence of a particular dominant type of foreground with a single color. The top three boxes in Fig. 7 plot the color-color diagrams for the three bands. There is a dominant color and trend line with a relatively small dispersion. In this discussion we define the color between any two bands, 1 and 2, as

$$\alpha_{12} = \frac{I_{\nu_1}}{I_{\nu_2}}. \tag{15}$$

Fig. 8 plots the average colors as a function of the intensity measured in a single band, and the error bars give the standard deviation for the same group of pixels. For brevity the results are shown only for map A in which no zodiacal model was subtracted. The horizontal axis is the deviation $\delta I_\nu$ of the single band in units of its standard deviation in this field. The solid line marks the average color of the entire field and the dotted lines delimit the $\pm 1$ standard deviation range in the colors for the entire field. The dispersion in their colors due to reddening or intrinsic color is very small ($< 10\%$ for all the fields), and there is a curious small trend in color with intensity that says that the brighter pixels also have higher color temperatures. Comparable results for the five fields and different $N_{cut}$ are summarized in Table 3. The first two columns show the threshold and the location of the field. Columns 3-8 give the mean values for the colors defined in Eq. (15), and the dispersion defined as $s_{12}^2 = \langle \alpha_{12}^2 \rangle - \langle \alpha_{12} \rangle^2$. Here and below $\langle ... \rangle$ denotes averaging over the ensemble of pixels in each field. The numbers in Table 3 are consistent with the near-infrared color-color diagrams for the DIRBE data discussed by Arendt et al. (1994). Since the Galactic sources, which are cataloged stars, and in Map A the zodiacal light, are so bright compared to known extragalactic sources, the colors we have identified are clearly those of the Galactic and zodiacal foreground emission.

We can use the colors $\alpha$ to determine what stellar population appears to be dominating the DIRBE $J$, $K$ and $L$-band emission. The typical mean values for the color indices shown in Table 3 are $\langle \alpha_{JK} \rangle = 1.40, \langle \alpha_{JL} \rangle = 1.85$ and $\langle \alpha_{KL} \rangle = 1.3$. The best stellar matches to these DIRBE colors appear to be K-M giant stars such as Alpha Bootis and Alpha Orionis (Arendt et al. 1994). This analysis suggests that the dominant contribution to the foreground color at high Galactic latitudes probably comes from K-M giants and (in map A) zodiacal light.

This empirical property that the near-IR foreground color has a small dispersion can be used to reduce the foreground contribution to $C(0)$ by taking a linear combination or scaled difference of two maps. On the other hand, Fig. 2 shows that most of the CIB in the near-IR is expected to come from galaxies whose spectra are significantly redshifted and is dominated by contributions from galaxies over a wide range of redshifts, and thus the CIB is not expected to be comparably attenuated by this procedure. These expectations motivate the color subtraction method we present below.

There are two *independent* contributions to the measured fluctuations in the linear combinations of the DIRBE maps: the one that comes from the *total* foreground (stars, zodiacal light, and cirrus), which we denote as $C_{*,\Delta}(0)$, and the one that comes from the extragalactic background which we label $C_{g,\Delta}(0)$. We use the $\Delta$ notation to indicate that these variances refer to the differences of maps at two different frequencies. The total $C_\Delta(0)$ is the sum of these two components. Let us assume that the Galactic and zodiacal foreground component has the mean color between any two frequencies 1 and 2 as defined by (14) of $S_{12} = \langle \alpha \rangle$ and the dispersion in the color is $s_{12} = \langle (\delta \alpha)^2 \rangle$. (This $S$ is different from the spectral energy distribution used in Eq. 8.) Assuming that the color fluctuations are uncorrelated with the intensities, the zero-lag signal due to the foreground in Band 1, $C_{*,1}(0)$, can be written as

$$C_{*,1}(0) \simeq (S_{12}^2 + s_{12}^2)C_{*,2}(0) + s_{12}^2 \langle I_{*,2} \rangle^2. \qquad (16)$$

Note that the approximations used in this equation are not required for the validity of the analysis of the maps, but are only to indicate that major improvements are achievable with color subtraction. (We will elaborate on the relevant mathematics in detail later in the context of the power spectrum formalism). Let us construct the quantity which depends on the parameters in two bands (we define $\delta_i \equiv I_i - \langle I_i \rangle$),

$$\Delta_{*,12} \equiv \delta_{*,1} - \beta \delta_{*,2}, \qquad (17)$$

where the quantity $\beta$ is for now an arbitrary number which will be quantified later. The zero-lag signal in the "color-subtracted" quantity defined above would be given by

$$C_{*,\Delta}(0) \equiv \langle \Delta_{*,12}^2 \rangle = [(S_{12} - \beta)^2 + s_{12}^2]C_{*,2}(0) + s_{12}^2 \langle I_{*,2} \rangle^2. \qquad (18)$$

It is clear from the above that if we were to choose $\beta$ close to the mean color for the foreground, $\beta \simeq S_{12}$, then the largest term in Eq. 18 could be made nearly zero and the contribution of the Galactic foreground confusion noise to the zero-lag signal $C_{*,\Delta}(0)$ in the difference map would be minimized. Furthermore, if $s \ll S$, that contribution can be decreased over the one specified in Eq. (16) by a fairly substantial amount.

Table 3 shows that indeed we have $s \ll S$. Thus the variance in the color-subtracted fluctuation $\Delta$ due to the confusion noise can be reduced by a substantial factor. We previously discussed that this contribution should reach a minimum when $\beta = S_{12}$. Fig. 9 shows the plot of $C_\Delta(0)$ for the data from all five fields as a function of $\beta$ for $J - K$, $J - L$ and $K - L$ subtraction, for map B and $N_{cut} = 3.5$. The minimum of $C_\Delta(0)$ is quite deep as it should be from Eq. (18). The reduction in the values of the zero-lag signal presented in Fig. 9 is comparable to that estimated in the end of Sec. 3.2 as a minimum requirement for finding the predicted CIB signal. Therefore, in constructing the color-subtracted quantities we use the values of $\beta$ for each field that correspond to the minimum of $C_\Delta(0)$. The minimum of $C_\Delta(0)$ is much more isotropic than the single band zero-lag signal, which in this figure would correspond to $\beta = 0$. The major anisotropy in the color subtracted signal comes from the SEP field. This field, however, contains the LMC galaxy and it is fair to exclude it from further discussion of the color subtracted signal. Similarly, the NEP field contains a significant fraction of pixels at $b^{II} < 25°$ and is therefore contaminated by the Galactic disk. The adjacent noncontaminated fields NEP1, SEP1 are plotted with thick solid and thick dashed lines in Fig. 9. As one can see the two new fields, SEP1, chosen to lie far away from the LMC, and NEP1, selected at larger $|b|$, have the same color subtracted signals as the other non-contaminated regions.

Fig. 10 shows the scatter diagram for the SGP field for the fluctuations in the color-subtracted quantities for B (upper panel) juxtaposed with those for single bands (lower panel). Map B has in general smaller fluctuations than Map A, showing that observational error has been reduced with more data and with the subtraction of the zodiacal light model. The scatter plots also show that a number of pixels are still quite far from the main distribution, suggesting that a non-Gaussian process has affected them. These plots do not show the core of the distribution, but the standard deviation is much less than that of the original maps. Whether the method is useful or not depends on how it affects the actual fluctuations of the extragalactic background.

### 4.5 Theoretical effects of color subtraction on predicted fluctuations

The color subtraction works well to suppress the fluctuations due to the Galactic foreground. We will now show that the predicted extragalactic background fluctuations are not removed by the color subtraction method because the redshifts give the background a different color than the foreground. The extragalactic signal would change by a different amount, and may even increase because of the combination of redshift and morphological effects. Using eq. (11) the zero-lag extragalactic signal due to (17) after convolving with the DIRBE beam as discussed in Sec. 2.3 would become:

$$C_\Delta(0) = 2.2 \times 10^{-19} Q_\nu^2 \int \frac{[E(z)W(\nu_1;\nu_2;z)\Psi(z)]^2}{(1+z)^2\sqrt{1+\Omega z}} [\frac{R_H(1+z)}{x(z)}]^{0.7} dz \ W^2 m^{-4} sr^{-2}, \qquad (19)$$

where

$$W(\nu_1;\nu_2;z) \equiv \frac{f_\nu(\nu_1(1+z);z) - \beta f_\nu(\nu_2(1+z);z)}{f_\nu(\nu_1;0)} = S(\nu_1;z) - \beta \frac{f_\nu(\nu_2;0)}{f_\nu(\nu_1;0)} S(\nu_2;z). \qquad (20)$$

Figs. 11a and 11b show the contributions of regions at different $z$ to the correlation function, for one band at a time and for the color-subtracted maps. These graphs show $\Psi^{-2}(z) z \partial C_\Delta(0)/\partial z$, the contribution to $C(0)$ as a function of $\log(z)$. As can be seen our method effectively reduces the contribution from the nearby galaxies, but because of the redshift effects on galaxies the overall zero-lag signal remains at the levels of the single band cases. In other words, the predicted color subtracted signal arises from galaxies in a window of redshifts centered on an early epoch.

Fig. 12 plots the theoretically computed amplitude (19) for the simple no-evolution models vs $\beta$ defined as above. The minimum value of the theoretically modelled zero-lag signal is decreased by a smaller factor compared to the single band number ($\beta = 0$) than in Fig. 9. (Note the different vertical scale in Fig. 11 compared to Fig. 9). This means that because the extragalactic fluctuations come from a wide range of redshifts, their colors have larger variance than the colors of the stars in our Galaxy (Table 3). These predicted numbers are subject to the same uncertainties and rescaling ($\propto$ (total flux)$^2$) as the single band $C(0)$. They also depend on the value of $\xi(r)$ at high redshift, which must be extrapolated from present data using a model.

## 5 Results: Measured Fluctuations

Table 4 shows the values of $\beta$ at which the minimum of $C_\Delta(0)$ occurs in the data for the five fields. The values of $\beta_{min}$ are roughly consistent in the various fields. We interpret the small differences as being due to different fractions of the stellar, zodiacal light, and cirrus contributions in the various fields. The variations of $\beta_{min}$ with the star clipping threshold and field are related to the differences in the relative contributions of zodiacal light (or the residuals after it is subtracted) and stars, and are consistent with the skewness trends on Fig. 5.

As shown above in sections 4.4 and 4.5, the foreground confusion noise in the linear combination signal could be decreased sufficiently to allow for detecting or setting strong constraints on the extragalactic background signal. Furthermore, as Fig. 9 shows for map B, the residual CIB correlation signal (when minimized with respect to $\beta$) now appears to be more isotropic, a necessary requirement for any signal of extragalactic origin. The major anisotropy is introduced by the SEP field, which contains the LMC, and where the single band fluctuations are significantly higher than in the other fields. The NEP field also appears to contribute to the anisotropy of $C_\Delta(0)$ although at a significantly lower level; as we discussed it is located close to the Galactic plane and is thus also contaminated. Due to the redshift effects, the value of the predicted $\beta_{min}$ for an extragalactic signal is shifted significantly compared to $\beta_{min}$ in Fig. 9. It is thus possible to reduce the foreground contribution to the measured $C(0)$ without greatly reducing the predicted contribution from the CIB. As Fig. 9 shows, at the values of $\beta$ corresponding to the predicted minimum of the extragalactic $C_\Delta(0)$ in Fig. 12, the data have a clearly anisotropic and therefore stellar signal.

Fig. 13 shows the distribution of the fluctuations in the color-subtracted data for the SGP field in Map B, which had the zodiacal light removed. Unlike the single band distributions shown in Fig. 5, the color subtracted quantities have symmetric distributions and are nearly Gaussian. The distribution of each $N_{cut} = 7$ clipped field has wide wings and can be represented by a sum of two roughly Gaussian distributions, one with small dispersion and over 90% of the pixels, and a second with a much larger dispersion. (It can also be represented as a sum of an inner Gaussian and some other-than-Gaussian second distribution responsible for the wide wings). Simulations show that this behavior is expected for the Poisson statistics of a foreground star distribution with a dispersion of individual colors. The narrower central Gaussian distribution arises from the majority of pixels that have many faint stars in them. For these pixels the color is an average over many stars and has a smaller dispersion than for the pixels that are dominated by single stars. The broader part of the distribution comes from the minority of pixels where a bright star is present. The narrower Gaussian has roughly the same dispersion for all the clipping thresholds. The contribution from the wider Gaussian is significantly decreased in the $N_{cut} = 5$ case, and completely disappears in the $N_{cut} = 3.5$ case, as expected from the simulations.

We emphasize that if the CIB has fluctuations with statistics like the stellar foreground, then by definition we can not recognize it by this test alone. However, in principle, they should not have the same statistics. First, the foreground stars are almost uncorrelated over small angular scales, while the extragalactic background is expected to have a particular dependence of correlation on angular separation. Second, we can use tables of the known stars to reduce their contributions, either by masking or by modeling, so the situation is not hopeless. As previously discussed, we compared the mask produced with the SAO bright star catalog with the masks produced with our algorithm, and they agree quite well. Third, the Poisson noise of stars should have a very different histogram than the background correlations. This can be seen from the discussion we gave of Eq. 6. The equivalent equation for foreground stars would include an integral factor of the form $\int x^{-4} x^2 dx$, where $x$ is the distance. This has a non-integrable singularity at $x = 0$, so that the foreground noise is dominated by the closest objects and depends on the cutoff of the integral at small $x$. This does not occur for the background fluctuations, in part because the clusters of galaxies at small $x$ are resolved by the DIRBE beam. The foreground galaxies are also suppressed by color subtraction.

Therefore, we believe that we are not suppressing significant amounts of extragalactic fluctuations with the clipping algorithm. Fig. 11 shows that the color subtraction method should suppress the foreground galaxies and clusters, since they have colors similar to our own Galaxy. Hence, the residual CIB fluctuations should come from large redshift, and should include many clusters in each beam. Therefore, these color-subtracted CIB fluctuations should be nearly Gaussian and should cover the entire sky, not just a small fraction of the pixels. The highly asymmetric histogram seen for foreground stars is due to the minority of stars close to the observer.

In Fig. 14 we replot the data to show the deviations from the Gaussian form. The figure plots the fraction of pixels versus the square of the deviation, normalized to the dispersion in the fields after the $N_{cut} = 7$ cut. In such plot a Gaussian distribution would be a straight line whose slope is inversely proportional to the variance. Thick lines correspond to the positive values of $\Delta$ (where we drop the subscript *,12) and thin ones to $\Delta < 0$. The distributions of the color subtracted quantities are very symmetric for all the cuts. The wider Gaussian distribution, which we identified with the fluctuations from individual bright stars of non-average colors in the $N_{cut} = 7$ case, contains a small fraction of pixels ($< 8 - 12\%$), and it disappears almost totally for smaller $N_{cut}$. We choose $N_{cut} = 3.5$ and remove these pixels of the wider part of the distribution in the computation of the final zero-lag signal.

Table 5 shows the values of the color subtracted $C(0)$ for the $N_{cut} = 3.5$, which as discussed above is a better estimate of the underlying zero-lag signal after removing the wider Gaussian population. We verified, and Fig. 13 shows this also, that the residual $C(0)$ computed in this way is approximately independent of the clipping threshold. As one can see from the table, the color subtracted signal is roughly isotropic. The degree of isotropy becomes higher if the two obviously contaminated fields, SEP containing most of the LMC, and NEP, partially located in the Galactic disk, are removed. Then the $J - K$ is isotropic to $\pm$ 18%, $J - L$ to $\pm$ 22%, and $K - L$ to $\pm$ 37%. Note that the rms deviation is the square root of $C(0)$ and is therefore isotropic to 9%, 11%, and 18% respectively. In all the cases the Lockman Hole gives either the minimum or next to the minimum value.

After all the above steps we again removed the residual linear gradients from the maps, but gradient removal makes a smaller difference for the color subtracted maps. This testifies to the efficiency of our color subtracting technique in removing foreground emission. Thus, this discussion suggests the following limits for the isotropic part of the color subtracted signal for these bands:

$$C_{J-K} < 7.3 \times 10^{-17} \, W^2 m^{-4} sr^{-2}, \tag{21}$$

$$C_{J-L} < 1.4 \times 10^{-16} \, W^2 m^{-4} sr^{-2}. \tag{22}$$

$$C_{K-L} < 1.3 \times 10^{-17} \, W^2 m^{-4} sr^{-2}. \tag{23}$$

The remaining anisotropies have not been explained but may well be due to differences in the foreground star populations and in the observational noise. We consider these measurements to be upper limits because we have not yet subtracted estimates for the residual noise expected from foreground stars, instrument noise, or attitude and pixelization errors.

We now want to know whether the remaining fluctuations measured in the color subtracted combinations are on the sky or in the observation process. Table 2 includes entries for the estimated contribution of observational noise to $C_\Delta(0)$ in the color subtracted maps, computed for the NGP and NEP regions with $N_{cut} = 7$ and 3.5, with the same values of $\beta$ used for that case. (The results for the other fields are similar). For $J - L$ and $K - L$ these noise levels are again significantly lower than values of $C_\Delta(0)$. For the $J - K$ case with $N_{cut} = 7$, the estimated noise contribution is only moderately smaller than the total $C_\Delta(0)$, and the comparison depends on the time scale for which the noise was estimated, while the noise contribution for $N_{cut} = 3.5$ is much smaller. This confirms our interpretation that for $N_{cut} = 7$ for $J - K$ the dominant observational noise is due to attitude errors, beam smearing, and pixelization effects. With more vigorous removal of the stars in single bands, we also remove the associated error sources. We have not subtracted the wider second Gaussian pixels from the computation of $C_\Delta(0)$ for the difference maps, but as we indicated before such a population is absent for $N_{cut} = 3.5$, and therefore the $N_{cut} = 3.5$ difference maps data are perhaps more indicative of the remaining observational noise.

Another test of the origin of the fluctuations seen in the difference maps is the shape of the spatial power spectrum. The power spectra of map B show clearly that the signals come from the sky, even for the color subtracted combinations. The signature is that the power spectra are much smaller at small scales, which can not be resolved by the DIRBE beam, than at scales of a degree or more, and the general shape is similar to the single-band power spectra before removing bright point sources. The difference maps C, D, and E all show spatial power spectra that are flat at angular scales less than $q^{-1} = 1°$, where $q$ is the spatial frequency in units of radians/degree, and give the true short term noise of the observations. The data are not sufficient to completely determine the origin of the sky fluctuations but it seems clear that they are not internal to the instrument. A full treatment of this topic is deferred to a further paper.

# 6  Limits on the CIB

We now proceed to the single band results on $C(0)$. As discussed above, the single-band measurements result in a very anisotropic zero-lag signal and can only be taken as upper limits on cosmological contributions. In order to interpret the results we choose the lowest values for $C(0)$ in Table 1:

$$C(0) < \begin{cases} 3.6 \times 10^{-16} \ W^2 m^{-4} sr^{-2} & in \ J, \\ 5.1 \times 10^{-17} \ W^2 m^{-4} sr^{-2} & in \ K, \\ 5.7 \times 10^{-18} \ W^2 m^{-4} sr^{-2} & in \ L. \end{cases} \quad (24)$$

The first two of these limits comes from the NGP field after the $N_{cut} = 3.5$ star removal; the $L$ band limit is from the LH region. It is interesting to note that the $K$ limit comes fairly close to the theoretical predictions of the diffuse background correlation function from Cole, Treyer and Silk (1992) based on modelling both deep $B$ and $K$ counts and invoking an evolving population of galaxies at high ($z > 1$) redshifts.

The model (evolution) independent limits implied by (14) for single band measurements would give:

$$(\nu I_\nu)_{z,rms} \leq \begin{cases} 200 \ nW m^{-2} sr^{-1} & in \ J, \\ 78 \ nW m^{-2} sr^{-1} & in \ K, \\ 26 \ nW m^{-2} sr^{-1} & in \ L. \end{cases} \quad (25)$$

These limits are independent of the spectral assumptions. In the $J, K, L$ bands they are significantly lower than direct (dc) estimates of the upper limits on the CIB derived from the darkest parts of DIRBE maps, and are comparable to the residuals in the fits of models (Hauser 1995, 1996).

Alternatively, one can interpret these limits in terms of evolutionary models. We will present here only a simple interpretation to illustrate the potential of the method and the results, which are probably accurate to within a factor of 2. (A more detailed interpretation will be presented in one of the forthcoming papers in this project). In order to interpret the results in terms of their theoretical significance we must rewrite eqs. (1) and (9) directly in terms of the background intensity $\nu I_\nu$ in band $\nu$, and the evolution parameters specified by (7), (8). Then the signals in the single bands can be written as

$$C(0) = 3.6 \times 10^{-3}(\nu I_\nu)^2 R_1^2(\nu, \Omega), \qquad (26)$$

where

$$R_1^2(\nu, \Omega) \equiv \frac{\int \frac{[E(z)S(\nu;z)\Psi(z)]^2}{(1+z)^2\sqrt{1+\Omega z}}[\frac{R_H(1+z)}{x(z)}]^\gamma dz}{[\int \frac{E(z)S(\nu;z)}{(1+z)^2\sqrt{1+\Omega z}}dz]^2}. \qquad (27)$$

In order to evaluate the limits on the CIB imposed by (25) we have to make some assumptions about the rate of the evolution of the clustering pattern (via $\Psi(z)$), the luminosity density (via $E(z)$) and the SED (via $S(\nu, z)$). The limits below are rather insensitive to these assumptions within the range discussed in the literature. For $\Psi(z)$ we consider the case of $\Psi(z) \propto (1+z)^{\gamma-2} \simeq (1+z)^{-1.7}$ and $\propto (1+z)^{-3}$, which describes a stable clustering pattern in either comoving or proper coordinates. For $E(z)$ we assume a power-law evolution $E(z) = (1+z)^P$ with $P = 0, 2$. Note that the power slope $P$ defined here includes both the evolution of the comoving number density (which determines when and how many galaxies form and die) *and* their luminosity (which determines how brightly they shine). We further assume at this stage of our project that the SED does not evolve; a more detailed interpretation of the results in terms of the evolution models will be presented at a later date. With these assumptions we find that for reasonable evolution parameters, the factor $R_1$ displays little variation with $\Psi(z), \Omega$, and $P$. For $0 \leq P \leq 2$, with $\Psi(z)$ varying in the above limits and $\Omega = 0.1, 1$ we find for the $J$ band that $3.6 \geq R_J \geq 3$. In the $K$ band the typical value is $R_K \simeq 2.0$ with equally small variations and in $L$ it is $R_L \simeq 1.4$. We use the measured single-band minimum values of $C(0)$ from (23) with these theoretical values of $R_1$ to obtain the following limits on the CIB:

$$\nu I_\nu < \begin{cases} 105(\frac{3}{R_J}) \, nW m^{-2} sr^{-1} & in \ J, \\ 60(\frac{2}{R_K}) \, nW m^{-2} sr^{-1} & in \ K, \\ 28(\frac{1.4}{R_L}) \, nW m^{-2} sr^{-1} & in \ L. \end{cases} \qquad (28)$$

These limits are again comparable to the residuals from Hauser (1995, 1996).

Similarly one can place limits on the CIB from the color-subtracted maps independently of evolution. For (21)-(23) eq. (14) leads to

$$\begin{cases} \langle [\nu_J(I_{\nu_J} - \alpha_{JK}I_{\nu_K})]\rangle_{z,rms} \leq 87 \, nW m^{-2} sr^{-1}, \\ \langle [\nu_J(I_{\nu_J} - \alpha_{JL}I_{\nu_L})]\rangle_{z,rms} \leq 129 \, nW m^{-2} sr^{-1}, \\ \langle [\nu_K(I_{\nu_K} - \alpha_{KL}I_{\nu_L})]\rangle_{z,rms} \leq 38 \, nW m^{-2} sr^{-1}, \end{cases} \qquad (29)$$

where $\alpha_{JK} = 1.5, \alpha_{JL} = 2.0, \alpha_{KL} = 1.75$. For the color subtracted quantities the equations (26, 27) can be modified in order to set upper limits on the mean fluxes, but it also turns out to be a significantly more model dependent procedure than for single bands. Here proper modelling would be required and if (21)-(23) could be treated as detections, this would allow the actual determination of a certain measure of the background fluxes. At this stage we postpone proper discussion of such a possibility, along with the relevant interpretation, to our forthcoming paper with a power spectrum analysis of the fields. Since the detector noise is small (Boggess *et al.* 1992), the possible sources for such fluctuations could be the long-sought CIB, the effects of patchy dust absorption, or most likely the confusion noise and residuals of the foreground star fluctuations. The power spectrum analysis is needed to show which of the above contributes to the detected signal.

# 7 Conclusions

We have presented the basic theory of fluctuations of the cosmic infrared background (CIB) and evaluated it for ranges of parameters suggested by direct observations of clustering, evolution, and spectral energy distributions of galaxies. We have developed algorithms for measuring these fluctuations in the DIRBE $J, K, L$ bands (1.25, 2.2, and 3.5 $\mu m$), by removing a model for the zodiacal light, masking out the obvious stars, and considering the residuals. We also showed that a suitable linear combination of maps at two different bands removes most of the residual fluctuations, a method we call color subtraction. The coefficients that remove these foreground fluctuations correspond to the colors of K-M giant stars, as appropriate for the main near-IR luminosity of the Galaxy.

The residual variances seen in the maps, evaluated as the angular correlation function at zero lag angle $C(0)$, are interpreted using the theory of fluctuations to obtain limits on the diffuse CIB due to clustered matter. These limits are comparable to theoretical expectations, and to the current residuals obtained from direct observations.

# 8 Acknowledgments

We thank the many members of the COBE data analysis staff who have spent the last several years reducing the DIRBE data to reliable maps. We particularly thank K. Mitchell and B. Franz, who developed the star blanking algorithm and read our manuscript carefully, Shoba Veeraraghavan and Eli Dwek for theoretical discussions, and W. Reach and T. Kelsall, who were particularly important in developing the zodiacal light model. Fruitful discussions with Jim Peebles are greatly appreciated. We also thank the engineering team, led by Donald Crosby and Loren Linstrom, who built the DIRBE instrument to very demanding specifications and made it work. Nearly 1500 people were required to complete the COBE. The COBE project was supported steadily by the NASA Astrophysics Division since studies were initiated in 1976. This work was partly supported through a grant from the NASA Long Term Space Astrophysics program. We thank the referee, Perry Hacking, for his careful reading and thoughtful comments.

# 9 References

Arendt, R. G., *et al.* 1994, Ap.J., **425**, L85.
Beichman, C.A. and Helou, G. 1991, Ap.J., **370**, L1.
Bernard, J.P. *et al.* 1994, A&A, **291**, L5.
Bernstein, G.M. 1994, Ap.J., **426**, 516.
Biller, S.D. *et al.* 1995, ApJ, **445**, 227.
Boggess, N.W. *et al.* 1992, Ap.J., **397**, 420.
Bond, J.R. *et al.* 1986, Ap.J., **306**, 428.
Bond, J.R. *et al.* 1991, Ap.J., **367**,420.
Boulanger, F., Baud, B. & van Albada, G. 1985, Astron. Ap., **144**, L9.
Broadhurst, T. *et al.* 1992, Nature, **355**, 55.
Bruzual, A.G. 1983, Ap.J., **273**, 105.
COBE 1995a, Skymap Information, avail. from National Space Sciences Data Center, or http://www.gsfc.nasa.gov/astro/cobe/skymap_info.html
COBE 1995b, DIRBE Explanatory Supplement, eds. M.G. Hauser, T. Kelsall, D. Leisawitz, and J. Weiland, National Space Sciences Data Center, available by anonymous FTP from nssdca.gsfc.nasa.gov.
Cole, S., Treyer, M. and Silk, J. 1992, Ap.J., **385**, 9.
Cole, S. *et al.* 1994, MNRAS, **267**, 541.
Chan, F.K. and O'Neill 1975, Computer Sciences Corporation EPRF Report 2-75.
Chan, F.K. and Laubscher, R.E. 1976, Computer Sciences Corporation EPRF Report 3-76.
Cowie, L. *et al.* 1990, Ap.J., **360**, L1.
Cowie, L. *et al.* 1991, Nature, **354**, 460.
Cowie, L. *et al.* 1994, Ap.J., **434**, 114.
Dermott, S.F. *et al.* 1994, Nature, **369**, 719.
Djorgovski, S. *et al.* 1995, Ap.J., **438**, L13.
Driver, S.P. *et al.* 1995, Ap.J., **449**, L29.
Dwek, E., and Slavin, J. 1994, Ap.J., **436**, 696.
Efstathiou, G. and Rees, M.J. 1988, MNRAS, **230**, 5P.
Efstathiou, G., Ellis, R. and Peterson, B.A. 1983, MNRAS, **232**,431
Efstathiou, G. *et al.* 1991, Ap.J., **380**, L47.
Franceschini, A. *et al.* 1991, Ap.J.Suppl., **89**, 285.
Gardner, J.P. *et al.* 1993, Ap.J., **415**, L9.
Gregorich, D.T. *et al.* 1995, AJ, **110**, 259.
Groth, E. and Peebles, P.J.E. 1977, Ap.J., **217**, 385.
Gunn, J. 1965, Ph.D. Thesis, Caltech. (unpublished)
Hacking, P.B., Condon, J.J. & Houck, J.R. 1987, Ap.J., **316**, L15.
Hacking, P.B. and Soifer, B.T. 1991, Ap.J., **367**, L49.
Hauser, M. *et al.* , 1984, Ap.J. (Letters), **278**, L15.
Hauser, M. 1993, in "Back to the Galaxy," AIP Conf. Proc. **278**, eds. S. Holt and F. Verter, (AIP:NY), 201.
Hauser, M. 1995, to be publ., in Proc. IAU Symposium 168, "Examining the Big Bang and Diffuse Background Radiations," Kluwer, Dordrecht.

Hauser, M. 1996, in "Unveiling the Cosmic Infrared Background," Univ. Maryland, April 1995, AIP Conf. Proc. **348**, ed. E. Dwek, (AIP:NY), 11-24.
Hu, E.M. and Ridgway, S.E. 1994, A.J., **107**, 1303.
Johnson, H. 1966, ARAA, **4**, 193.
Kashlinsky, A. 1992, Ap.J., **399**, L1.
Kashlinsky, A. 1993, Ap.J., **406**, L1.
Kashlinsky, A. 1994, in "Evolution of the Universe and its Observational Quest," ed. K. Sato, p.181, Universal Academy Press, Inc. - Tokyo.
Kashlinsky, A. and Rees, M. 1983, MNRAS, **205**, 955.
Kashlinsky, A. and Jones, B.J.T. 1991, Nature, **349**, 753.
Koo, D. *et al.* 1993, Ap.J., **415**, L21.
Koo, D. and Kron, R. 1992, ARAA, **30**, 613.
Lilly, S. *et al.* 1991, Ap.J., **369**, 79.
Lilly, S. 1993, Ap.J., **411**, 501.
Limber, D. N. 1953, Ap.J., **117**, 134.
Lockman, F.J., Jahoda, K. and McCammon, D. 1986, Ap.J., **354**, 184.
Lonsdale, C.J. *et al.* 1990, Ap.J., **358**, 60.
Loveday, S. *et al.* 1992, Ap.J., **390**, 338.
Low, F.J. *et al.* 1984, Ap.J., **228**, L19.
Maddox, S. *et al.* 1990, MNRAS, **242**, 43p.
Martin, C. and Bowyer, S. 1989, Ap.J, **338**, 677.
Matsumoto, T. *et al.* 1988, Ap.J., **332**, 575.
Melott, A. 1992, Ap.J.,**393**, L45.
Neuschaefer, L.W. and Windhorst, R.A. 1995, Ap.J., **439**, 14.
Pagel, B.E.J. 1993, in "The Cold Universe," eds. Montmerle *et al.* , Editions Frontieres, p.345
Peebles, P.J.E. 1980, "Large Scale Structure of the Universe," Princeton Univ. Press.
Peebles, P.J.E. 1987a, Nature, **327**, 210.
Peebles, P.J.E. 1987b, Ap.J., **315**, L73.
Picard, A. 1991, Ap.J., **368**, L7.
Pritchet, C.J. and Infante, L. 1992, Ap.J., **399**, L35.
Rakos, K.D. and Schombert, J.M. 1995, Ap.J., **439**, 47.
Reach, W. *et al.* , 1995, Nature, **374**, 521.
Reach, W. *et al.* , 1996, in "Unveiling the Cosmic Infrared Background," AIP Conf. Proc. **348**, Ed. E. Dwek, (AIP: New York), 37-46.
Saunders, W. *et al.* 1991, Nature, **349**, 32.
Saunders, W., Rowan-Robinson, M., and Lawrence, A. 1992, MNRAS, **258**, 134.
Schechter, S. 1976, Ap.J, **203**, 297.
Shectman, S. 1973, Ap.J., **179**, 681.
Shectman, S. 1974, Ap.J., **188**, 233.
Silverberg, R.F. *et al.* , 1993, Proc. SPIE Conf. **2019**, "Infrared Spaceborne Remote Sensing," ed. M.S. Scholl (Bellingham: SPIE), 180.
Smoot, G. *et al.* 1992, Ap.J., **396**, L1.
Soifer, B.T., Neugebauer, G. and Houck, J.R. 1987, ARAA, **25**, 187.
Stecker,, F., and DeJager, O. 1993, Ap.J., **415**, L71.
Sykes, M.V. and Walker, R.G. 1992, Icarus, **95**, 180.

Tyson, A. 1988, A.J., **96**, 1.
Wainscoat, R.J. *et al.* 1992, Ap.J.Suppl., **83**, 111.
Wang, B. 1991, Ap.J.,**374**,465.
Weiland, J.L. *et al.* , in "Unveiling the Cosmic Infrared Background," AIP Conf. Proc. **348**, Ed. E. Dwek, (AIP: New York), 74-80.
White, R.A., and Mather, J.C. 1991, "Databases from the Cosmic Background Explorer (COBE)," Databases & On-line Data in Astronomy. Astrophysics and Space Science Library, eds. M.A. Albrecht and D. Egret, (Dordrecht: Kluwer), 171, pp. 30-34.
Wright, E.L. 1992, in "The Infrared and Sub-mm Sky After COBE," M. Signore & C. Dupraz, eds. (Kluwer:Netherlands), 231.
Yoshii, Y. and Peterson, B.A. 1991, Ap.J., **372**, 8.
Yoshii, Y. and Takahara, F. 1988, Ap.J., **326**, 1.
Yoshii, Y. 1993, Ap.J., **403**, 552.

Table 1. $C(0)$ for various cutoff thresholds in each band.

| Region | $N_{cut}$ | $N_{total}$ | $C(0)$ (W$^2$m$^{-4}$sr$^{-2}$) | | |
|---|---|---|---|---|---|
| | | | $J$ | $K$ | $L$ |
| NEP  | 7 | 956   | $2.6 \times 10^{-14}$  | $4.9 \times 10^{-15}$  | $5.1 \times 10^{-16}$ |
| NEP1 | 7 | 902   | $2.8 \times 10^{-14}$  | $5.9 \times 10^{-15}$  | $6.1 \times 10^{-16}$ |
| NGP  | 7 | 910   | $1.1 \times 10^{-14}$  | $1.8 \times 10^{-15}$  | $1.8 \times 10^{-16}$ |
| SEP  | 7 | 1,000 | $28.4 \times 10^{-14}$ | $62.1 \times 10^{-15}$ | $77.1 \times 10^{-16}$ |
| SEP1 | 7 | 955   | $3.5 \times 10^{-14}$  | $6.65 \times 10^{-15}$ | $6.9 \times 10^{-16}$ |
| SGP  | 7 | 960   | $1.3 \times 10^{-14}$  | $2.3 \times 10^{-15}$  | $2.5 \times 10^{-16}$ |
| LH   | 7 | 247   | $6.0 \times 10^{-14}$  | $1.4 \times 10^{-14}$  | $13.8 \times 10^{-16}$ |
| NEP  | 5 | 822   | $11.3 \times 10^{-15}$ | $23.5 \times 10^{-16}$ | $25.2 \times 10^{-17}$ |
| NEP1 | 5 | 712   | $5.9 \times 10^{-15}$  | $9.2 \times 10^{-16}$  | $8.95 \times 10^{-17}$ |
| NGP  | 5 | 669   | $1.85 \times 10^{-15}$ | $2.9 \times 10^{-16}$  | $3.0 \times 10^{-17}$ |
| SEP  | 5 | 757   | $34.6 \times 10^{-15}$ | $75.5 \times 10^{-16}$ | $91.0 \times 10^{-17}$ |
| SEP1 | 5 | 833   | $13.4 \times 10^{-15}$ | $27.1 \times 10^{-16}$ | $28.2 \times 10^{-17}$ |
| SGP  | 5 | 719   | $2.3 \times 10^{-15}$  | $3.9 \times 10^{-16}$  | $3.8 \times 10^{-17}$ |
| LH   | 5 | 174   | $4.1 \times 10^{-15}$  | $6.9 \times 10^{-16}$  | $6.9 \times 10^{-17}$ |
| NEP  | 3.5 | 327 | $21.0 \times 10^{-16}$ | $36.3 \times 10^{-17}$ | $37.3 \times 10^{-18}$ |
| NEP1 | 3.5 | 210 | $12.3 \times 10^{-16}$ | $20. \times 10^{-17}$  | $19.7 \times 10^{-18}$ |
| NGP  | 3.5 | 269 | $3.6 \times 10^{-16}$  | $5.1 \times 10^{-17}$  | $5.8 \times 10^{-18}$ |
| SEP  | 3.5 | 302 | $83. \times 10^{-16}$  | $172. \times 10^{-17}$ | $192. \times 10^{-18}$ |
| SEP1 | 3.5 | 902 | $9. \times 10^{-16}$   | $15.9 \times 10^{-17}$ | $17.1 \times 10^{-18}$ |
| SGP  | 3.5 | 273 | $6.2 \times 10^{-16}$  | $8.1 \times 10^{-17}$  | $8.1 \times 10^{-18}$ |
| LH   | 3.5 | 57  | $4.8 \times 10^{-16}$  | $6.5 \times 10^{-17}$  | $7.3 \times 10^{-18}$ |

Table 2. Noise levels for DIRBE average and difference maps, $N_{cut} = 7, 3.5$.

| Band | Region | $N_{weeks}$ | $\beta$ | C7 | C35 | D7 | D35 | E7 | E35 |
|---|---|---|---|---|---|---|---|---|---|
| J | NEP | 40 | 0 | 10.4 | 1.4 | 10.9 | 1.5 | 3.4 | 1.4 |
| K | | | 0 | 3.2 | 0.3 | 2.8 | 0.3 | 2.4 | 0.25 |
| L | | | 0 | 0.6 | 0.09 | 0.3 | 0.11 | 0.3 | 0.09 |
| J-K | | | 1.35 | 5.7 | 3.3 | 6.2 | 3.1 | 9.8 | 2.8 |
| J-L | | | 2.4 | 9.4 | 4.2 | 6.6 | 5.2 | 10. | 3.8 |
| K-L | | | 1.95 | 1.7 | 1.0 | 1.8 | 1.4 | 1.3 | 1.0 |
| J | NGP | 19 | 0 | 10.9 | 3.8 | 11. | 3.6 | 2.4 | 1.0 |
| K | | | 0 | 1.9 | 1.0 | 1.7 | 0.8 | 0.65 | 0.4 |
| L | | | 0 | 0.4 | 0.3 | 0.4 | 0.3 | 0.14 | 0.14 |
| J-K | | | 1.35 | 5.5 | 3.9 | 5.5 | 3.9 | 4.1 | 2.8 |
| J-L | | | 2.4 | 9.5 | 10. | 9.0 | 8.7 | 7.2 | 7.4 |
| K-L | | | 1.95 | 2.4 | 2.4 | 2.2 | 2.1 | 1.8 | 1.8 |

Units: $10^{-17}$ W$^2$m$^{-4}$sr$^{-2}$. All maps have zodiacal light removed. Map C column is estimate of noise contribution to map B, derived from 8 weeks added with alternating + and - signs, and assuming that the variance scales as 1/(number of weeks). Map D is like Map C but comparing 4 weeks in first half of mission with 4 weeks in last half. Map E is from comparing 2 successive weeks. The same value of $\beta$ was used for all the color-subtracted maps.

Table 3. Color parameters for map B (41 weeks after zodi subtraction).

| Region | $N_{cut}$ | $\langle \alpha_{JK} \rangle$ | $s_{JK}$ | $\langle \alpha_{JL} \rangle$ | $s_{JL}$ | $\langle \alpha_{KL} \rangle$ | $s_{KL}$ |
|---|---|---|---|---|---|---|---|
| NEP | 7 | 1.43 | 0.12 | 2.49 | 0.24 | 1.74 | 0.09 |
| | 5 | 1.44 | 0.10 | 2.48 | 0.19 | 1.72 | 0.08 |
| | 3.5 | 1.49 | 0.06 | 2.49 | 0.13 | 1.68 | 0.08 |
| NGP | 7 | 1.50 | 0.15 | 1.91 | 0.32 | 1.28 | 0.19 |
| | 5 | 1.50 | 0.12 | 1.81 | 0.22 | 1.21 | 0.15 |
| | 3.5 | 1.51 | 0.11 | 1.69 | 0.15 | 1.12 | 0.10 |
| SEP | 7 | 1.33 | 0.11 | 2.35 | 0.21 | 1.76 | 0.09 |
| | 5 | 1.36 | 0.09 | 2.38 | 0.17 | 1.75 | 0.08 |
| | 3.5 | 1.40 | 0.06 | 2.40 | 0.12 | 1.72 | 0.06 |
| SGP | 7 | 1.45 | 0.14 | 1.86 | 0.30 | 1.29 | 0.18 |
| | 5 | 1.45 | 0.11 | 1.78 | 0.21 | 1.23 | 0.14 |
| | 3.5 | 1.48 | 0.09 | 1.69 | 0.18 | 1.14 | 0.11 |
| LH | 7 | 1.46 | 0.15 | 2.20 | 0.23 | 1.51 | 0.19 |
| | 5 | 1.53 | 0.11 | 2.17 | 0.21 | 1.43 | 0.14 |
| | 3.5 | 1.56 | 0.08 | 2.08 | 0.15 | 1.33 | 0.08 |

Table 4. The values of $\beta_{min}$ for map B.

| Region | $J-K$ | | | $J-L$ | | | $K-L$ | | |
|---|---|---|---|---|---|---|---|---|---|
| $N_{cut}$ | 7 | 5 | 3.5 | 7 | 5 | 3.5 | 7 | 5 | 3.5 |
| NEP | 1.25 | 1.20 | 1.35 | 2.40 | 2.30 | 2.65 | 1.95 | 1.90 | 2.05 |
| NGP | 1.30 | 1.40 | 1.35 | 2.60 | 2.60 | 2.00 | 1.95 | 1.85 | 1.30 |
| SEP | 1.20 | 1.25 | 1.25 | 2.10 | 2.20 | 2.30 | 1.75 | 1.75 | 1.80 |
| SGP | 1.30 | 1.35 | 1.50 | 2.40 | 2.55 | 2.50 | 1.90 | 1.85 | 1.55 |
| LH | 1.15 | 1.35 | 1.45 | 2.30 | 2.60 | 2.35 | 1.95 | 1.95 | 1.65 |

Table 5. Correlations for $N_{cut} = 3.5$

| Region | $C(0)$ W$^2$m$^{-4}$sr$^{-2}$ | | |
|---|---|---|---|
| | J–K | J–L | K–L |
| NEP | $18. \times 10^{-17}$ | $31. \times 10^{-17}$ | $2.2 \times 10^{-17}$ |
| NEP1 | $9.2 \times 10^{-17}$ | $19. \times 10^{-17}$ | $1.9 \times 10^{-17}$ |
| NGP | $8.8 \times 10^{-17}$ | $14. \times 10^{-17}$ | $2.1 \times 10^{-17}$ |
| SEP | $35. \times 10^{-17}$ | $66. \times 10^{-17}$ | $4.8 \times 10^{-17}$ |
| SEP1 | $7.45 \times 10^{-17}$ | $15. \times 10^{-17}$ | $1.6 \times 10^{-17}$ |
| SGP | $10.5 \times 10^{-17}$ | $22. \times 10^{-17}$ | $2.8 \times 10^{-17}$ |
| LH | $7.3 \times 10^{-17}$ | $16. \times 10^{-17}$ | $1.3 \times 10^{-17}$ |

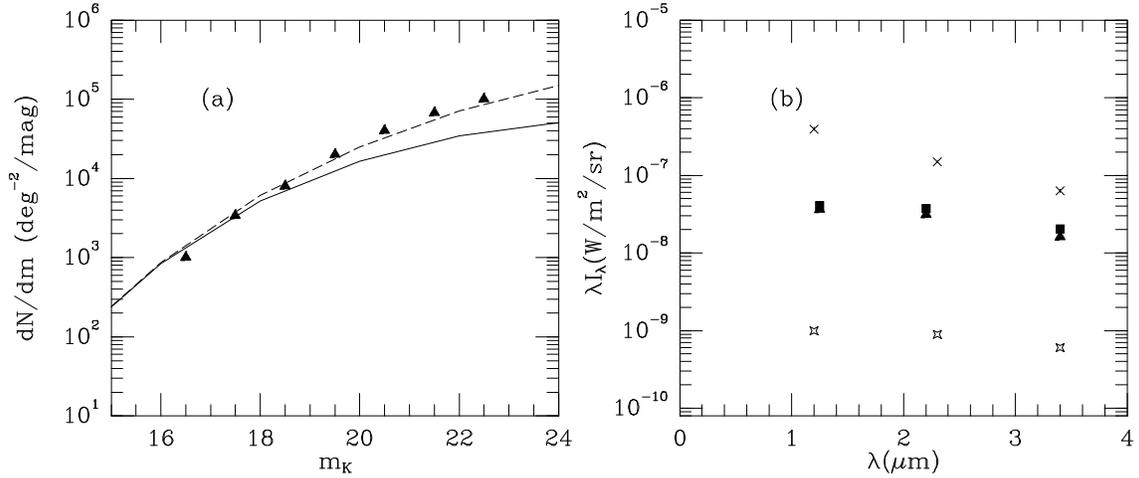

Figure 1: (a) K-band faint galaxy counts from Cowie *et al.* (1991) are plotted with triangles. Solid line shows the no-evolution prediction for the $\Omega = 1$ Universe and the dashed line shows the no-evolution prediction for $\Omega = 0.1$. (b) Crosses plot the current DIRBE upper limits on the diffuse infrared background from Hauser (1995). Filled squares and triangles correspond to the predicted background from galaxies as described in the text; squares are for $\Omega = 1$ and triangles for $\Omega = 0.1$. 4-pointed stars show the DIRBE detector noise levels in these bands from Boggess *et al.* (1992). Actual map uncertainties are dominated by confusion noise from stars.

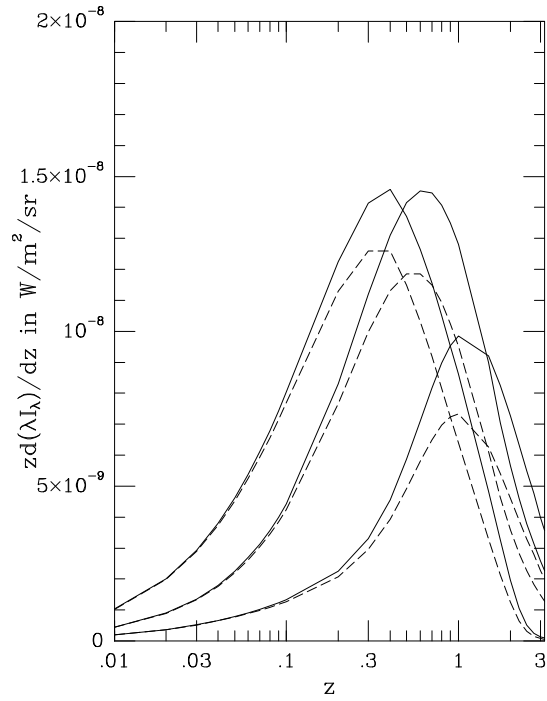

Figure 2: The redshift distribution of $\nu I_\nu = \lambda I_\lambda$ is plotted for the no-evolution models for the $J, K, L$ bands. Solid lines correspond to $\Omega = 1$ and the dashed lines to $\Omega = 0.1$.

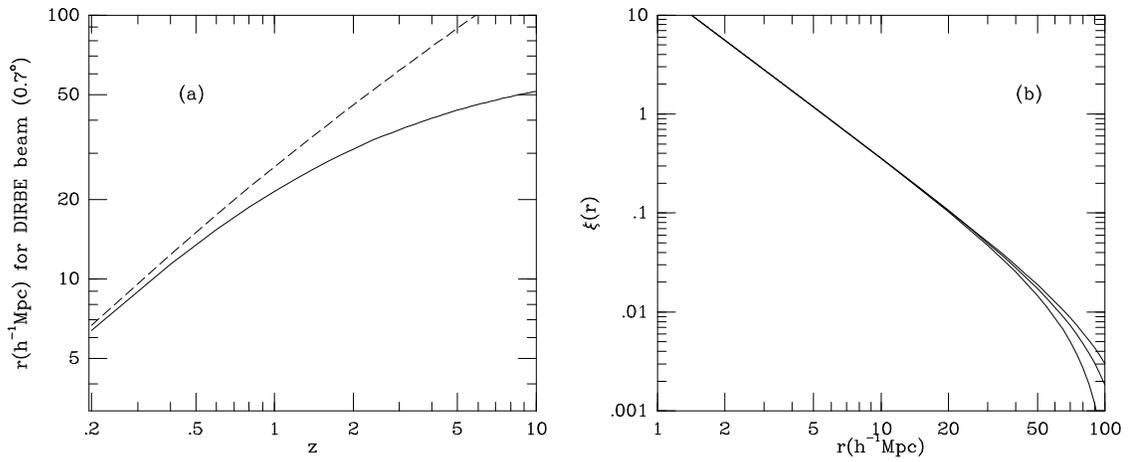

Figure 3: (a) The comoving scale subtended by the DIRBE beam (0.7°) is plotted vs the redshift. Solid line corresponds to $\Omega = 1$; the dashed line corresponds to $\Omega = 0.1$. (b) Present-day correlation function for the APM data normalized to the COBE DMR (Kashlinsky 1992). The three lines correspond to the wave numbers for transition to the Harrison-Zeldovich regime of $k_0^{-1} = 40$, 50, $60h^{-1}$Mpc respectively. $k_0 \geq 50h^{-1}$Mpc is required to match APM data and COBE results.

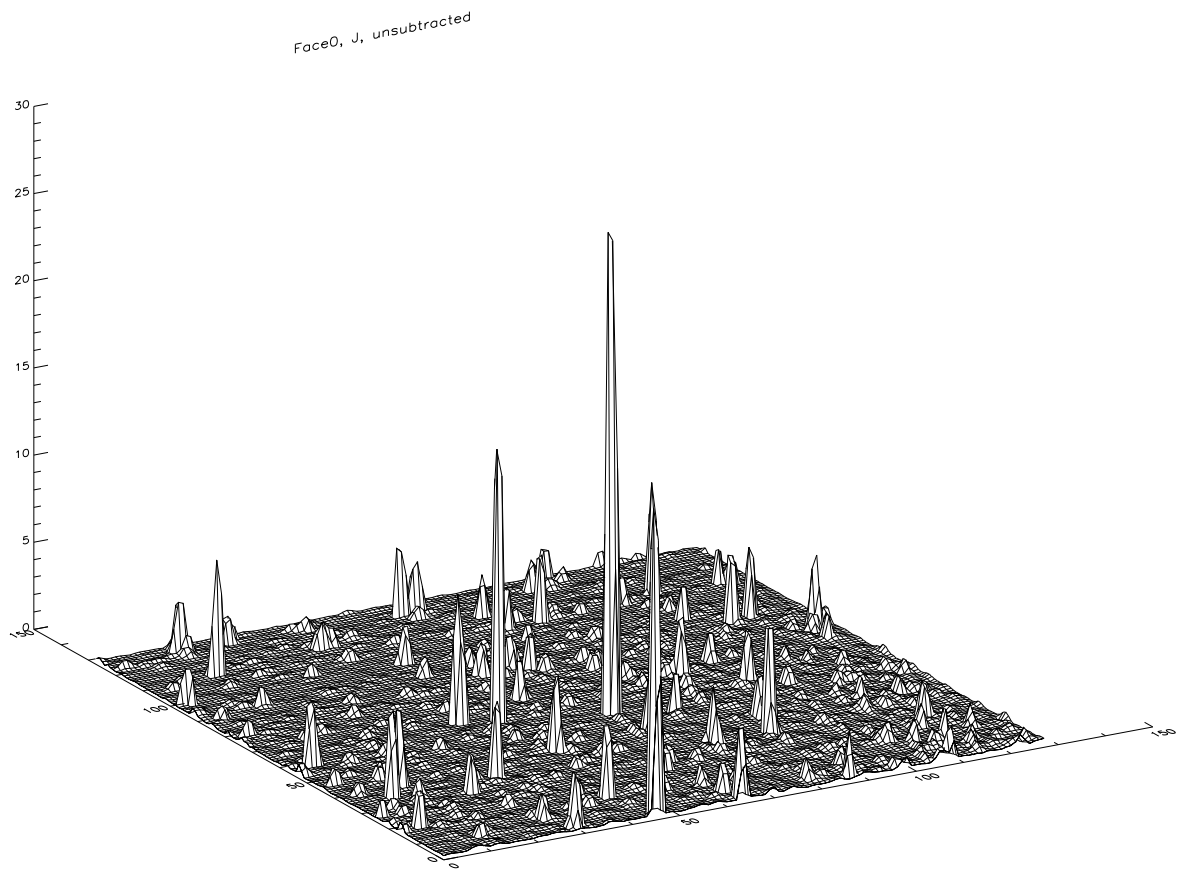

Figure 4: Surface plot of *J*-band intensities in MJy/sr for the field in cube Face 0 for Map A.

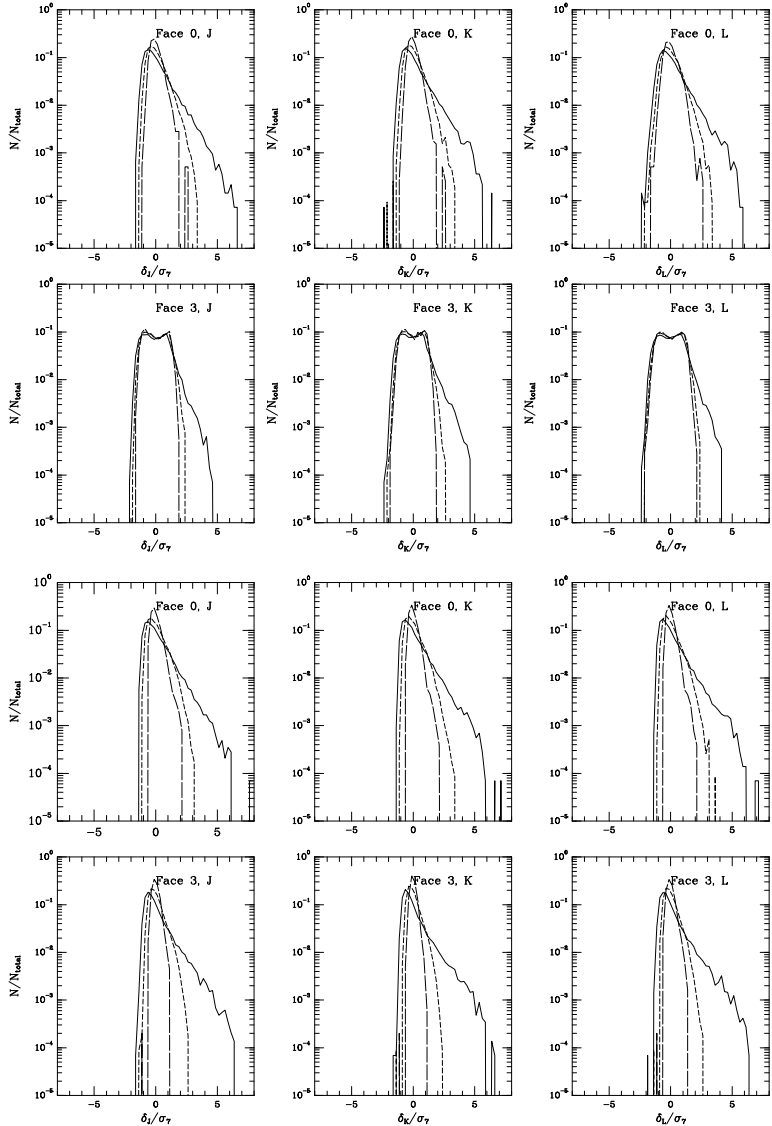

Figure 5: Histograms of the pixels remaining in the $J, K, L$ bands after the star/peak subtraction for fields in cube Face 0 and 3. Solid lines: $N_{cut} = 7$. Short dashes: $N_{cut} = 5$. Long dashes: $N_{cut} = 3.5$. The horizontal axis is the deviation of intensity measured in units of $\sigma_7$; the standard deviation of the residuals for that field/frequency with $N_{cut} = 7$. Note the asymmetry and the tail on the positive side of $\delta$ for $N_{cut} = 7$. The steep left sides of the distributions occur because of the statistics of discrete stars. Top group of six histograms correspond to map A (4 weeks of data without the zodiacal subtraction performed). Note the differences between the fields in cube Face 0 (near ecliptic pole) and Face 3 (in ecliptic plane) which contain different contributions from the zodiacal light. Histograms are shown of the pixels remaining in the $J, K, L$ bands after the star/peak subtraction are shown for the fields in cube Face 0 and 3. Solid lines correspond to $N_{cut} = 7$, short dashes to $N_{cut} = 5$, and the long dashes are for $N_{cut} = 3.5$. The horizontal axis for each small plot is the deviation of intensity measured in units of $\sigma_7$, the standard deviation of the residuals of the map for that field and frequency made with $N_{cut} = 7$. Bottom group of six histograms for Map B (41 weeks and zodiacal light removed). Note now the similarities between the two fields and their similarity to the of the zodiacal light removal.

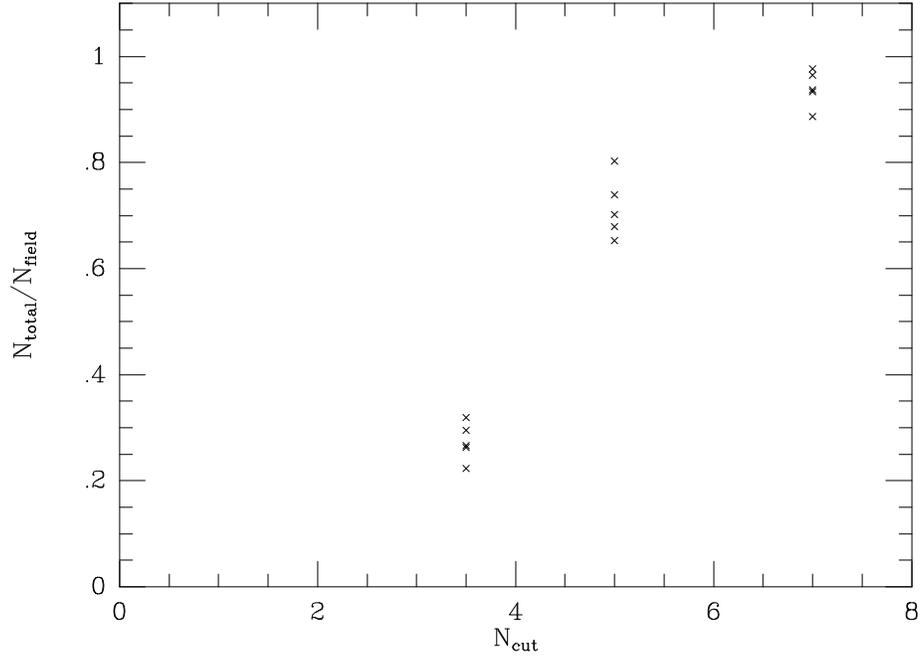

Figure 6: The fraction of the initial pixels ($N_{field}$) left in the five fields in map B plotted vs the clipping threshold $N_{cut}$.

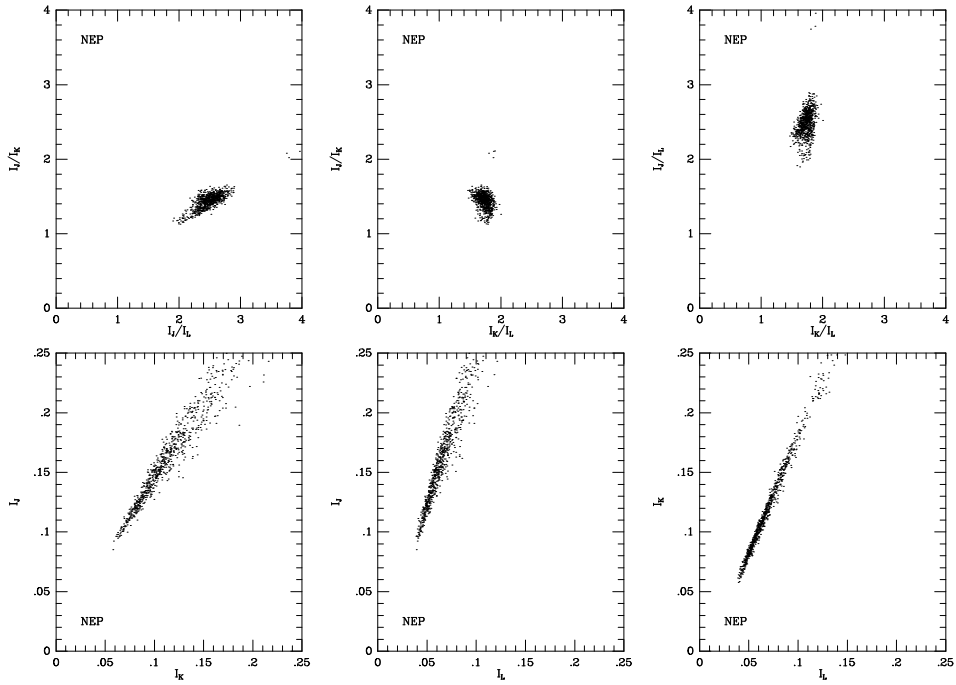

Figure 7: Scatter diagram for the NEP field after the subtraction of stars with the $N_{cut} = 5$ threshold for map A. (see text). Fluxes are plotted in MJy/sr.

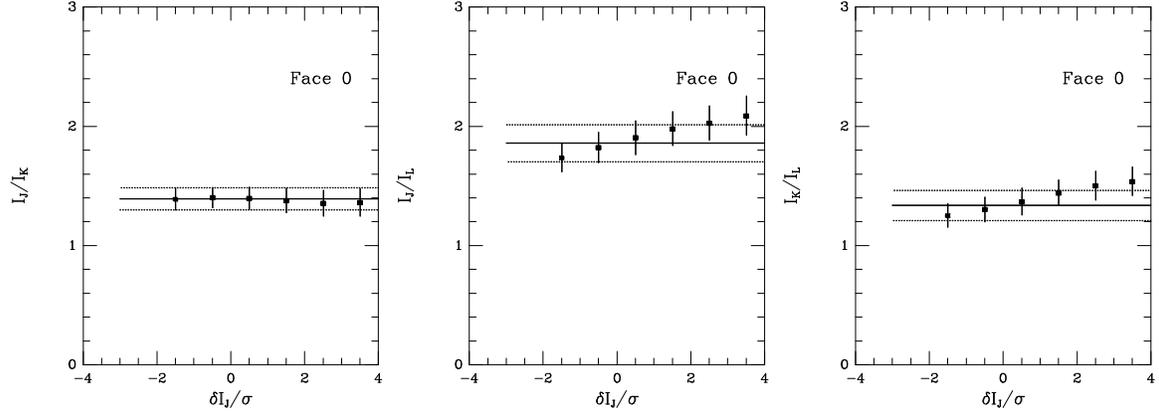

Figure 8: Colors of the pixels plotted for the $J$-band in Fig. 5 are shown as squares with their error bars. Dashed line is the average color for the field and the $\pm 1\sigma$ levels around it are shown with dotted lines. For brevity the plots are shown for map A.

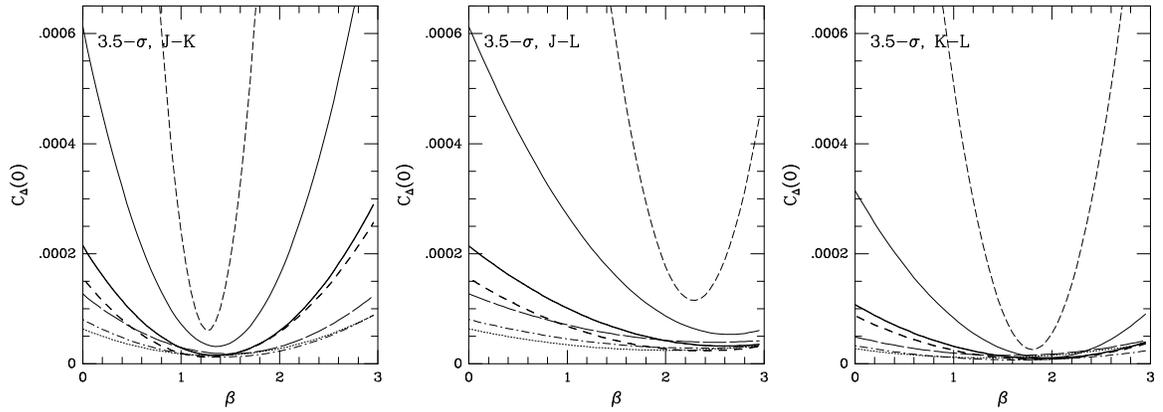

Figure 9: $C_\Delta(0)$ in units of $MJy^2 sr^{-2}$ is plotted vs $\beta$ defined in eq. (16) (see text). The plot is drawn for map B for $N_{cut} = 3.5$. Solid lines correspond to the NEP field, dotted to NGP, short dashes to SEP (which contains the LMC), the long dashes to the SGP and the dot-dashed lines to the LH field. Thick solid lines corresponds to NEP1 and thick short dashed line to SEP1.

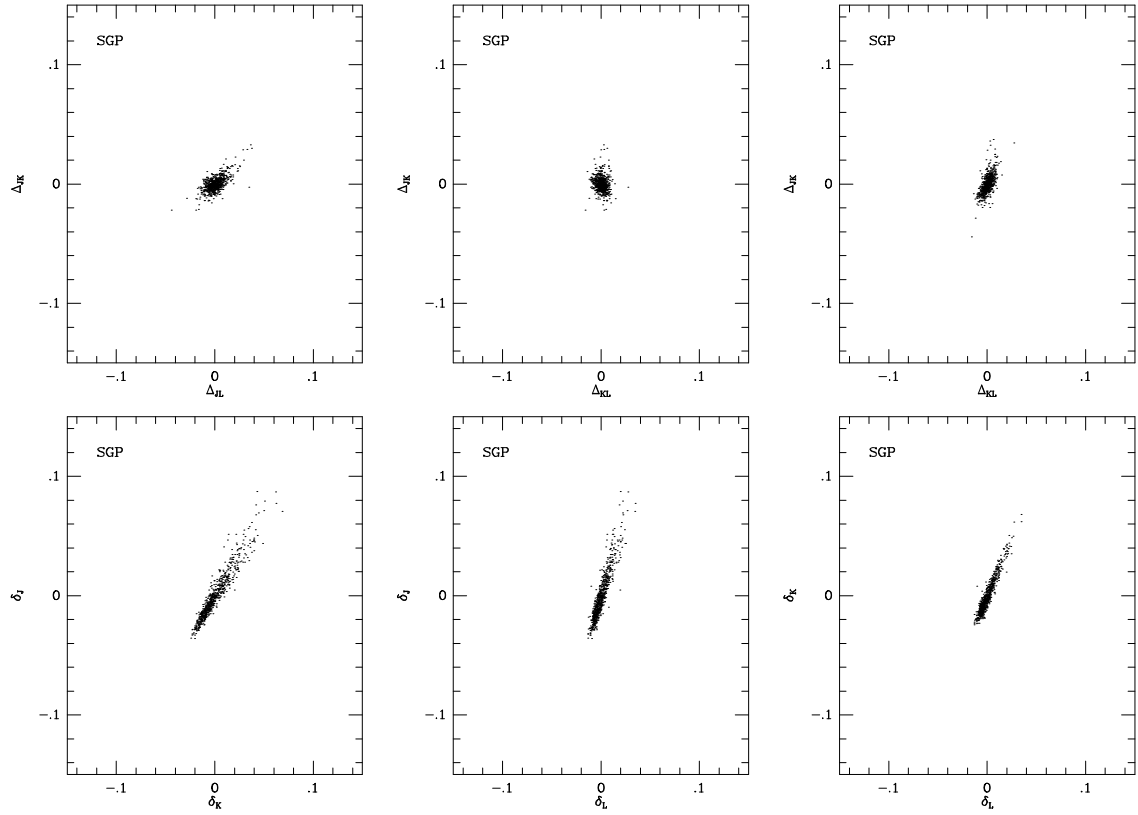

Figure 10: Scatter diagram for single band fluctuations (bottom three boxes) in units of MJy/sr is juxtaposed with that of the color subtracted fluctuations with $\beta = \langle \alpha \rangle$ (top three panels) for map B.

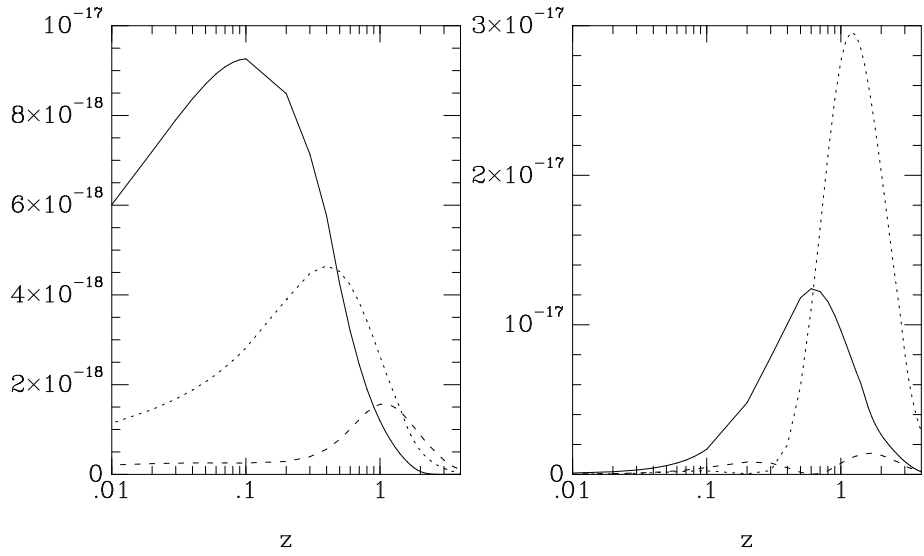

Figure 11: Differential contribution to the overall variance, $\Psi^{-2}(z)zdC(0)/dz$ is plotted vs the redshift $z$. (a) single band (eq. 10) for the no-evolution model. Solid line corresponds to $J$, dotted to $K$ and the dashed line to $L$ bands. (b) same as (a) only for the two-band subtraction: $J - K$ is denoted with solid line, $J - L$ with dotted and $K - L$ subtraction is plotted as dashed line. The colors used are those given in Table 2.

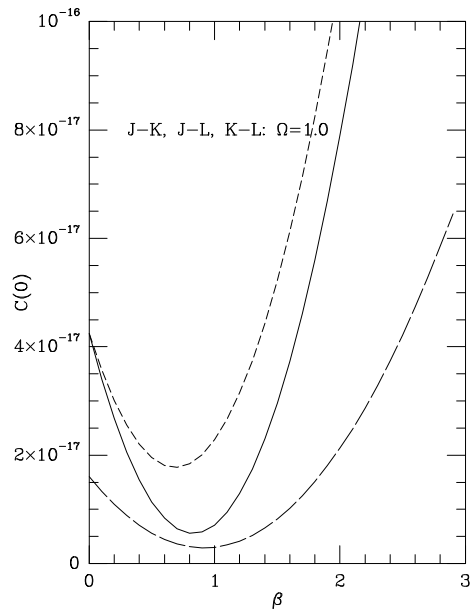

Figure 12: Theoretical value of $C_\Delta(0)$ vs $\beta$ computed for no-evolution models from eq. (15). Solid line corresponds to $J - K$, short dashes to $J - L$ and long dashes to $K - L$. The numbers plotted are for $\Omega = 1$; for $\Omega = 0.1$ they would be marginally higher.

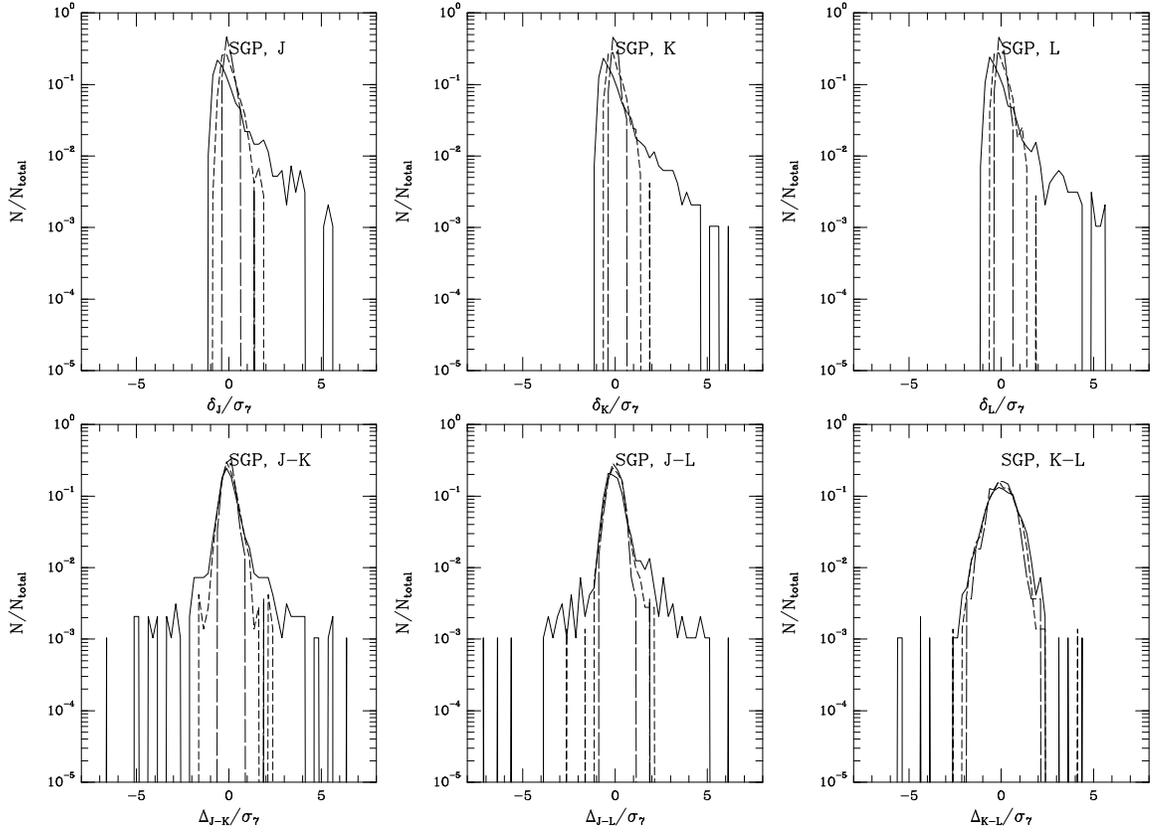

Figure 13: Distribution of the pixel fluctuations for the color-subtracted quantities for $\beta = \beta_{min}$. Same line notation as in Fig. 5. Note the symmetry of the distributions and the clear tails described by a second wide distribution for $N_{cut} = 7$ (solid lines). Map B data are used in the plots.

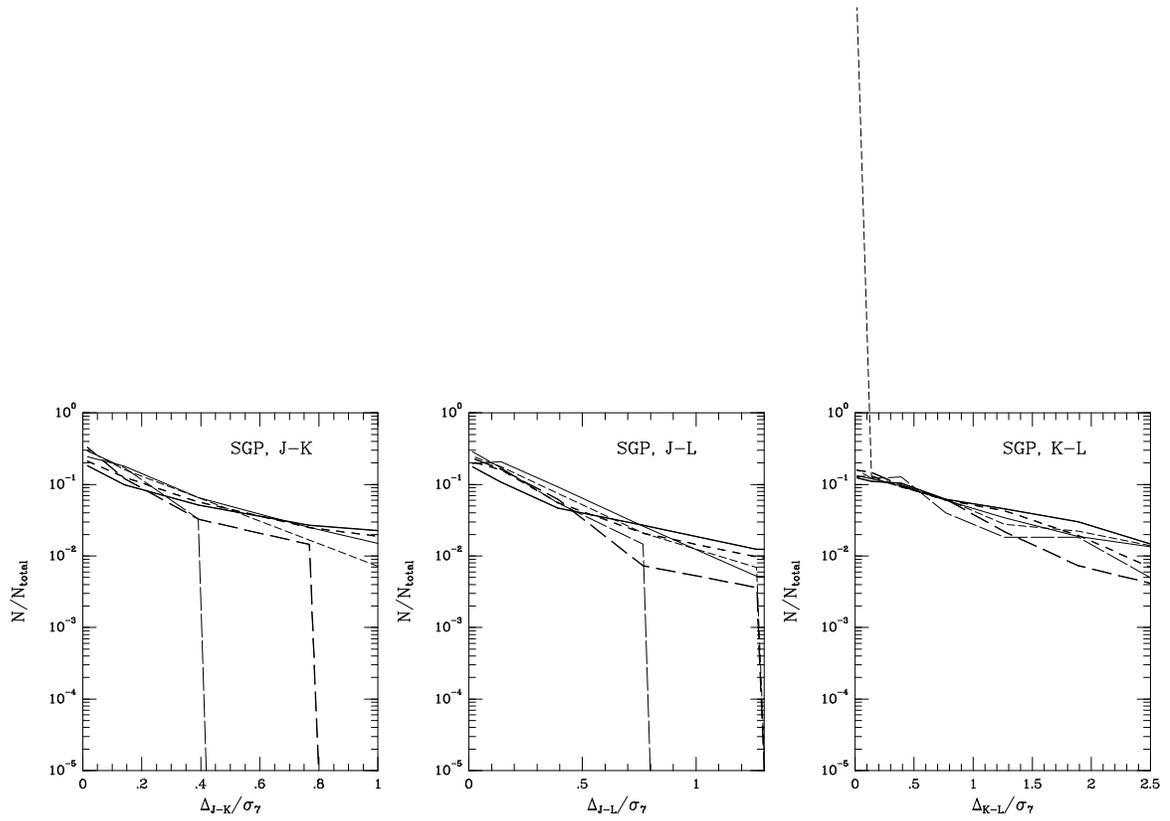

Figure 14: Distribution of the fluctuations is plotted for the SGP field for map B. Thick lines correspond to $\Delta > 0$ and thin to $\Delta < 0$; otherwise line notation is the same as in Figs. 5 and 12. The data are plotted in such a way that the Gaussian distribution would be a straight line with slope determined by the dispersion.